\documentclass[amsmath,amssymb,aps,prx,twocolumn,showpacs]{revtex4-2}
\usepackage{graphicx}
\usepackage{bm}
\usepackage{hyperref}
\usepackage{color}
\usepackage{footmisc}

\begin{document}

\title{The Jaynes-Cummings model breaks down when the cavity significantly reduces the emitter's free-space emission rate}

\author{Martin Blaha}
\author{Arno Rauschenbeutel}
\author{J{\"u}rgen Volz}
\email{juergen.volz@hu-berlin.de}
\affiliation{Department of Physics, Humboldt-Universit\"at zu Berlin, 12489 Berlin, Germany.}

\date{\today}

\newcommand{\mb}[1]{\textcolor{red}{#1}}

\begin{abstract}
Strong coupling between a single resonator mode and a single quantum emitter is key to a plethora of experiments and applications in quantum science and technology and is commonly described by means of the Jaynes-Cummings model. Here, we show that the Jaynes-Cummings model only applies when the cavity does not significantly change the emitter's emission rate into free-space. Most notably, the predictions made by the Jaynes-Cummings model become increasingly wrong when approaching the ideal  emitter-resonator systems with no free-space decay channels. We present a Hamiltonian that provides, within the validity range of the rotating wave approximation, a correct theoretical description that applies to all regimes. As minimizing the coupling to free-space modes is paramount for many cavity-based applications, a correct description of strong light-matter interaction is therefore crucial for developing and optimizing quantum protocols.  
\end{abstract}
\maketitle

Light-matter interaction is at the foundation of quantum optics and forms the basis for many applications in quantum science and technology. At the fundamental quantum level, this interaction is described by the Jaynes-Cummings Hamiltonian \cite{jaynes_comparison_1963}. The latter describes the coupling between a two-level quantum emitter and a single, quantized mode of the electromagnetic field. Light-matter interaction in such a cavity quantum electrodynamics (CQED) setting \cite{walther2006} has been experimentally realized with numerous resonator types ranging from bulk cavities in the optical \cite{Ye1999,Pinkse2000} or microwave domain \cite{brune_quantum_1996} to integrated nanophotonic and plasmonic resonators in the optical domain
\cite{Aoki2006,Thompson2013,junge2013,santhosh2016vacuum, hugall2018plasmonic, bitton2020vacuum} to on-chip superconducting coplanar waveguide microwave resonators \cite{schoelkopf2008wiring,Hu2012}. Depending on the resonator type, emitters ranging from atoms to molecules to quantum dots to defect centers, to superconducting qubits have been coupled \cite{wang_2019,michler2000quantum,yoshie_vacuum_2004,Merkel2020,zhang2018strongly,fink_2008}.
A common goal of most studies is to maximize the coupling strength between the resonator mode and the emitter, in order to enter the regime of strong-coupling or, equivalently, high cooperativity. It is the regime that maximizes the performance of CQED systems for many applications \cite{reiserer_cavity-based_2015}.

Here, we show that the Jaynes-Cummings (JC) model in general does not provide a valid description of CQED systems when the fraction of free-space emission of the excited emitter becomes too \emph{small}. Indeed, surprisingly, the JC description is only valid when the solid angle covered by the cavity mode is small. Otherwise, the emitter may induce a significant change of the cavity field in a single roundtrip. This violates the implicit assumption made by the JC model that a global cavity field prevails that can be described by a single operator. Consequently, when the fraction of free-space emission is too small, the JC model makes qualitatively and quantitatively wrong predictions.

To gain a quantitative understanding of why and when this happens, we start with the JC Hamiltonian in the rotating-wave approximation
\begin{equation}
\hat{H}_{\textrm{JC}}/\hbar = \omega_0 \hat{\sigma}^+ \hat{\sigma}^- + \omega_a \hat{a}^{\dagger} \hat{a} + g\left( \hat{a}^{\dagger} \hat{\sigma}^- + \hat{a} \hat{\sigma}^+\right)+\hat U_\text{cpl}~,
\label{eq:JC}
\end{equation}
where $\hat{\sigma}^+$ (resp. $\hat{\sigma}^-$) is the operator that creates (resp. annihilates) an excitation in the emitter and $\hat{a}^{\dagger}$ ($\hat{a}$) creates (annihilates) a photon in the cavity. The frequencies $\omega_0$ and $\omega_a$ denote the emitter and cavity resonance, respectively, and $g$  the emitter-photon coupling strength, which we assume to be real. In order to take coupling to the environment and probing of the system into account, we add the operator $\hat U_\text{cpl}$, which takes the form
\begin{eqnarray}
\hat{U}_\text{cpl}& =&-i\gamma_l \hat{\sigma}^+ \hat{\sigma}^- -i\kappa_l \hat{a}^{\dagger} \hat{a}+\hat U_\text{probe}~.\label{eq:Ucoupl}
\end{eqnarray}
Here, the first two terms describe loss from the system due to scattering by the emitter and dissipation in the resonator with the dipole decay rate, $\gamma_l$, and field decay rate, $\kappa_l$, respectively. The last term allows to externally drive the system and is given by Eq.~(\ref{eq:Uprobe}) in Appendix \ref{app:JC}.
For negligible coupling to the environment, i.e., $\gamma_l$ \& $\kappa_l\rightarrow0$, and without drive, $\hat U_\text{probe} =0$, one is left with the well-known Jaynes-Cummings Hamiltonian \cite{jaynes_comparison_1963}. 
\begin{figure}[tb]
	\includegraphics[width=0.8\columnwidth]{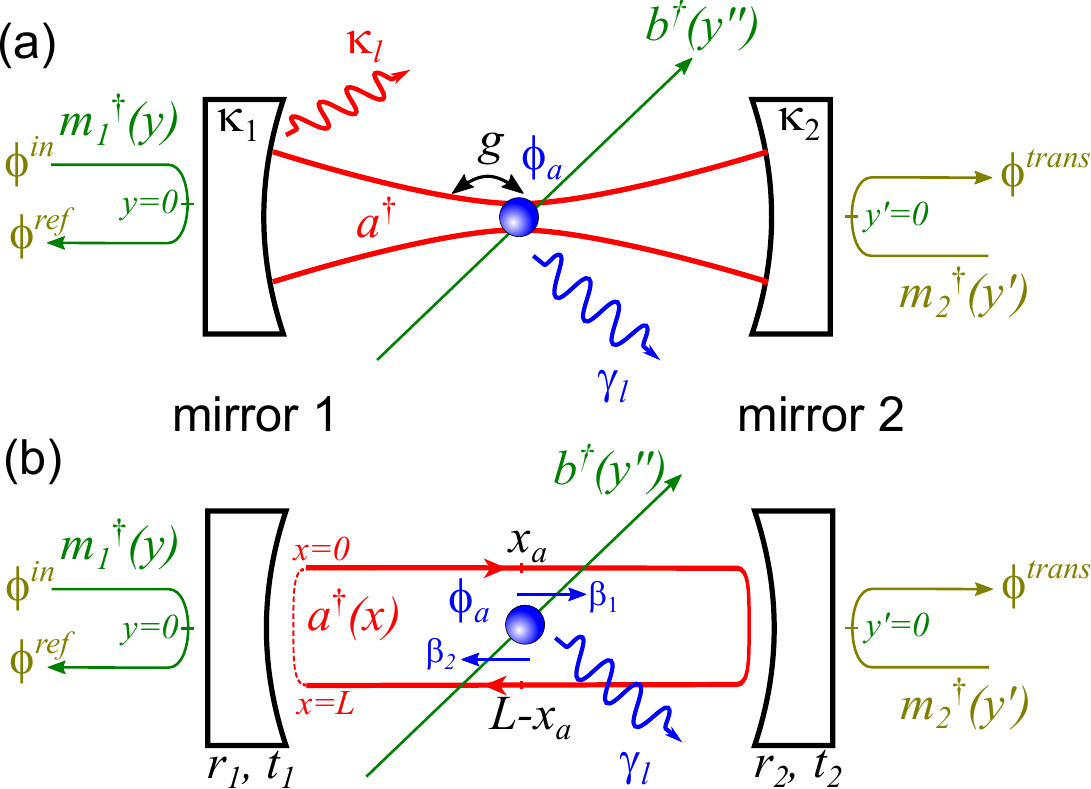}	
	\caption{Schematic representation of the parameters used in the two models. For both cases, the two cavity mirrors and the emitter are probed using the external propagating fields $\hat m^\dagger_1$, $\hat m^\dagger_2$ and $\hat b^\dagger$, respectively. Note that the axis for these fields are chosen such that the resonator mirror or the emitter is located at the respective zero-position. (a) For the Jaynes-Cummings Hamiltonian, the global cavity field $\hat a^\dagger$ couples to the fields incident on mirror $i$ with the in-coupling rate $\kappa_i$ and to the emitter with the emitter-resonator coupling strength $g$.  Losses are described by the cavity loss rate $\kappa_l$ and the emitter decay rate into the environment $\gamma_l$.	(b) For our cascaded model, the cavity field is represented by a position-dependent field $\hat a^\dagger(x)$ with $x\in[0,L]$ that circulates in the cavity and interacts with the emitter at the positions $x_a$ and $L-x_a$ with a strength given by $\beta_1$ and $\beta_2$, respectively. Coupling of the cavity field to the outside modes at the mirrors is described by the beamsplitter-type matrices that relate the external and internal fields via the amplitude reflection and transmission coefficients $r_i$ and $t_i$ of the mirror.}
	\label{fig:scheme}
\end{figure}
Now, we consider the situation of a Fabry-P\'erot resonator, which is constantly fed with a probe field via the incoupling mirror. In the steady state, the time-independent Schr\"odinger equation based on Hamiltonian~(\ref{eq:JC}) can be analytically solved in the low excitation regime, see App.~\ref{app:JC}. The emitter-induced amplitude decay rate of the cavity mode is given by
\begin{eqnarray}
\Gamma=\frac{\rho_{e}}{\rho_a}\gamma_l&=&\frac{g^2}{\gamma_l}~,
\end{eqnarray}
see App.~\ref{app:lossrate}.
Here, $\rho_e$ is the excited state population of the emitter and $\rho_a$ the probability of finding a photon in the cavity mode.
Surprisingly, $\Gamma$ diverges when approaching an ideal emitter--resonator system with no decay of the emitter into the environment, i.e., for $\gamma_l \rightarrow0$. Thus, when $\gamma_l$ becomes too small, the rate of change of the cavity field, $ \Gamma $, will exceed the inverse of the roundtrip time of the resonator field, $\tau_c^{-1}=c/L=\nu_{\text{fsr}}$. In other words, the resonator field changes significantly during one cavity round trip. In this case, the resonator field can no longer be described by a single operator $\hat a$, and the JC model no longer describes the system correctly. Here, $L$ and $\nu_{\text{fsr}}$ are the roundtrip optical path length and the free spectral range of the resonator, respectively, and $c$ denotes the speed of light.
A similar argument can be made using the characteristic reaction time of the emitter, see App.~\ref{app:lossrate}. 

We note that the situation we describe here is different from other limits where the JC model breaks down, i.e., the super- and ultrastrong coupling regimes, which are reached when the coupling strength $g$ becomes comparable to $2\pi\nu_{\text{fsr}}$ or $\omega_0$,
respectively \cite{meiser_superstrong_2006,Sundaresan2015,yoshihara_superconducting_2017,johnson_observation_2019,kockum2019,Ultrastrong_2019}. In particular, the limit we discuss here only depends on the fraction of free-space decay channels of the emitter and can be reached for arbitrarily small coupling strengths.

In order to obtain a more intuitive break-down condition of the JC model, we consider the emission with rate $\gamma$ into all modes in the absence of cavity enhancement and separate it into two parts: the fraction that overlaps with the cavity mode (channeling efficiency $\beta$) and the remaining part ($1-\beta$). With the cavity present, the emission rate of an excited emitter into the environment thus reads $\gamma_l=(1-\beta)\gamma$.

For a Fabry-P\'erot resonator, the emitter-resonator coupling strength $g$ can be expressed in terms of this channeling efficiency $\beta$ as $g=2\sqrt{\beta\gamma\nu_{\text{fsr}}}$ [Appendix \ref{app:chiral}] where we assumed that the emitter is located at an anti-node of the cavity standing wave. Using this link between $\beta$ and $g$, we can reformulate the condition $\Gamma\ll\nu_{\text{fsr}}$ in terms of $\beta$ and obtain
\begin{equation}
\beta\ll\frac{1}{5} \label{eq:beta}
\end{equation}
as the condition for the validity of the JC model. This means, when the cavity mode encompasses 1/5 or more of all possible free-space modes, the basic assumption that the resonator field does not change significantly during one photon roundtrip in the cavity does not hold anymore. Consequently, one expects significant deviations between model and reality. Note that even when  exceeding this boundary the rotating-wave approximation still holds. 
For standard mirror cavities the boundary in Eq.~(\ref{eq:beta}) is reached when the waist of the cavity mode approaches the optical wavelength $w_0\approx\lambda$, i.e., $\beta\approx0.15$ [Appendix \ref{app:waist}], which has already been experimentally realized in the optical \cite{Benedikter2017} and microwave regime \cite{gleyzes2007quantum}. While it is difficult to significantly further increase $\beta$ in conventional cavities, $\beta$ factors exceeding the boundary in Eq.~(\ref{eq:beta}) can be reached when interfacing the emitter with photonic \cite{Fam_decay,Skoff2018}, plasmonic \cite{santhosh2016vacuum, hugall2018plasmonic, bitton2020vacuum} or superconducting structures \cite{fink_2008}. In particular with microwave circuit QED systems values of $\beta\approx1$ have already been experimentally realized \cite{Sundaresan2015,bosman2017multi,Kuzmin2019}. Thus, for these experimental platforms a new description of CQED or circuit CQED is required.

In order to obtain a theoretical description that is valid for arbitrary $\beta$, we use an ansatz for the Hamiltonian, where we explicitly describe the cavity field as position-dependent, i.e., we replace the field operators $\hat a$ and $\hat a^\dagger$ by their position dependent counterparts $\hat a(x)$ and $\hat a^\dagger(x)$. Following the waveguide description outlined in \cite{Shen2009,Blaha2021}, in our model we treat the resonator mode as a propagating field and derive the emitter-cavity Hamiltonian for a Fabry-P\'erot resonator as
\begin{eqnarray}
&&\hat{H}/\hbar=\omega_0\hat\sigma^+\hat\sigma^-\nonumber\\
&&-ic\int\limits_{0}^{L/2}dx \hat a^\dagger(x)\partial_x \hat a(x)-ic\int\limits_{L/2}^{L}dx' \hat a^\dagger(x')\partial_{x'} \hat a(x')\nonumber\\ 
&&+V_{a_1}(\hat a^\dagger(x_a)\hat \sigma^-+\hat a(x_a)\hat \sigma^+)\nonumber\\
&&+V_{a_2}(\hat a^\dagger(L-x_a)\hat \sigma^-+\hat a(L-x_a)\hat \sigma^+)\nonumber\\
&&+c\left[\hat U_{m_1}(L,0)+\hat U_{m_2}(L/2,L/2)\right]+\hat{U}_\text{cpl}. 
\label{eq:ourJC}
\end{eqnarray}
In this expression $\partial_x=(i\omega_\text{lin}/c+\partial/\partial x)$ where $\omega_\text{lin}$ is the frequency around which the dispersion relation of the guided light is approximately linear and $c$ is the corresponding group velocity of the light \footnote{For fields propagating in a dispersionless medium the expression simplifies to $\partial_x=\partial/\partial x$}. The operator
\begin{eqnarray}
\hat{U}_{m_k}(x,x')&=& -ir_k\hat a(x)\hat a^\dagger(x')-t_k\hat a(x)\hat m_k^\dagger(0)\nonumber\\
&&-ir_k \hat m_k'(0)\hat m_k^\dagger(0)-t_k \hat m_k'(0)\hat a^\dagger(x')
\end{eqnarray}
is the beamsplitter matrix that describes the cavity mirror $k$ with the field transmission and reflection coefficients $t_k$ and $r_k$, respectively, that, without loss of generality, we assume to be positive. Here, $\hat m_k(y)$ ($\hat m^\dagger_k(y)$) is the photon annihilation (creation) operator for light fields incident and reflected from cavity mirror $k$, positioned at $y=0$ with respect to the external mode, see Fig \ref{fig:scheme}.

\begin{figure}[tb]
	\includegraphics[width=0.9\columnwidth]{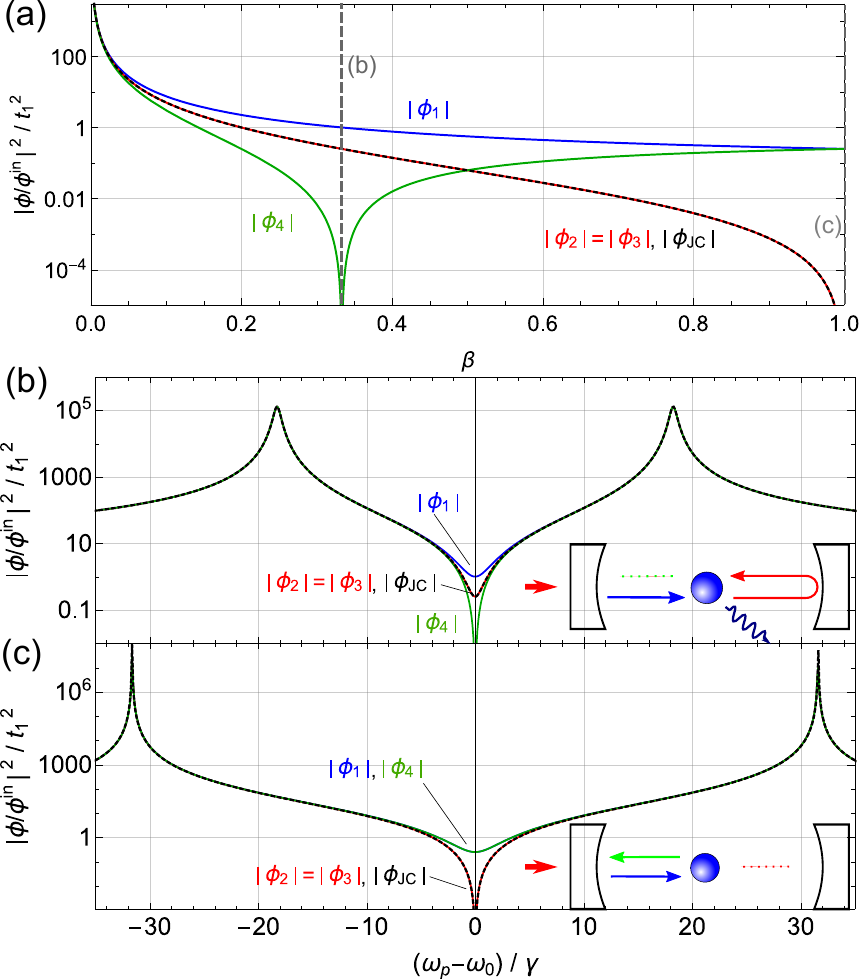}	
	\caption{(a) On-resonance field strength in the cavity as a function of atom channeling efficiency $\beta$ for a high-finesse resonator for an atom positioned at an anti-node of the cavity field mode ($\alpha=\pi$). The solid lines indicate the prediction of our cascaded model and the dashed line that of the JC model. (b) and (c) Excitation of the different cavity fields as a function of light-resonator detuning for $\beta=1/3$ and $\beta=1$, respectively, assuming $t_1^2=t_2^2=1-r_1^2=1-r_2^2=10^{-4}$ and $\nu_{\text{fsr} }=250\gamma$.
	}
	\label{fig:beta1}
\end{figure}

Equation (\ref{eq:ourJC}) is equivalent to the Jaynes-Cummings Hamiltonian in Eq.~(\ref{eq:JC}) but with a position-dependent running-wave resonator field that circulates in the cavity between the mirrors located at position $x=0$ and $x=L/2$, see Fig.~\ref{fig:scheme}. For a Fabry-P\'erot resonator, the light will thus interact twice with the emitter at the position $x_a$ and $L-x_a$ with the coupling strength $V_{a_1}$ and $V_{a_2}$, respectively. From this, it is obvious that, in general, in the steady state there cannot be one well-defined photon field in the cavity but, e.g., for the Fabry-P\'erot resonator, there are four regions in the cavity with four different amplitudes $\phi_i$ of the circulating wave, see insert in Fig.~\ref{fig:beta1}(a).

As we here consider a position-dependent resonator field, photon loss from the cavity cannot simply be described by a cavity loss rate $\kappa_l$ but instead, one has to explicitly take into account at which position $x$ the loss occurs. The Hamiltonian description in Eq.~(\ref{eq:ourJC}) does not account for cavity losses. However, for the case where losses only occur upon reflection on the cavity mirrors the former can straightforwardly be included using non-unitary beamsplitter matrices ($r_k^2+t_k^2\neq1$) for the mirrors. The term $\hat{U}_\text{cpl}$ describes the interaction of the resonator and the emitter with external modes [Eq.~(\ref{eq:Ucpl}) in Appendix \ref{app:cascHam}]. 
For the case, where we only consider probing the resonator via the cavity mirrors and are not interested in the evolution of photons scattered into freely-propagating modes, we can summarize the effect of the latter by introducing the complex energy term $-i\gamma_l\hat\sigma^+\hat\sigma^-$ and the coupling Hamiltonian becomes
\begin{eqnarray}
&&\hat{U}_\text{cpl}=-i\gamma_l\hat \sigma^+\hat \sigma^-\nonumber\\
&&-ic\sum\limits_{k=1}^2\Big[ 
\int\limits_{-\infty}^0dy\;\hat m_k'^\dagger(y)\partial_y \hat m'_k(y)+\int\limits_0^{\infty}dy\;\hat m_k^\dagger(y)\partial_y \hat m_k(y)\Big].\nonumber\\
\end{eqnarray}
In the following, we assume the most common situation of symmetric, i.e., direction independent emitter-resonator mode coupling where $\beta$ is the total decay probability of an excited emitter into both propagation directions of the resonator mode, i.e., $V_{a_1}=V_{a_2}=\sqrt{\beta\gamma c}=g\sqrt{L}/2$, where $g$ is the coupling strength for an emitter at the antinode of the cavity field. In Appendix~\ref{app:chiral} we also discuss the fully chiral case, $ V_{a_2} = 0 $ \cite{junge2013,lodahl_chiral_2017}, where the emitter couples only to one propagation direction of the circulating light field, a scenario realized in experiments like \cite{scheucher_quantum_2016, tang_2019}.

For a quantitative comparison of the JC and our cascaded model, we now analyze the models' predictions for the steady state for the case of a symmetric emitter-mode coupling and a Fabry-P\'erot resonator that is weakly driven by a field with frequency $\omega_p$ that is incident on mirror 1. The full analytic solutions of our cascaded model are given in Eqs.~(\ref{eq:our_solution1})-(\ref{eq:our_solution4}) in Appendix~\ref{app:FP}. 
On resonance ($\omega_0=\omega_a=\omega_p$), for high-finesse optical cavities ($r_1,r_2\approx 1$) and channeling efficiencies that are not too small, $\beta \sin^2(\alpha_0/2)>t_1,t_2$, the solutions simplify to
\begin{eqnarray}
\left.\begin{array} {rrr}
\phi_1\\
\phi_2\\
\phi_3\\
\phi_4
\end{array}
\right\}&=&\phi_{in}\cdot\frac{-it_1}{4\beta\sin^2(\alpha_0/2)}\left\{ 
\begin{array}{ll}
(\beta e^{i\alpha_0}-1)\\
(\beta-1)\\
(1-\beta)\\
(\beta(e^{-i\alpha_0}-2)+1) \\
\end{array}\right.\label{eq:phi_res}
\end{eqnarray}
where $\phi_{in}$ is the amplitude of the probe light field and the parameter $\alpha_0$ indicates the emitter's position along the optical axis in the resonator standing wave mode, with $\alpha_0=0$ ($\alpha_0=\pi$) corresponding to the node (anti-node) [Eq.~(\ref{eq:alpha}) in Appendix~\ref{app:FP}]. The four field regions $\phi_1...\phi_4$ are indicated in Fig.~\ref{fig:beta1}(a). For the JC model the steady state solutions predict [Eq.~(\ref{eq:compareJC1}) in Appendix~\ref{app:JC_solution}]
\begin{equation}
\phi_{\textrm{JC}}=\frac{it_1}{4\beta\sin^2(\alpha_0/2)}\phi_{in}\cdot(\beta-1).
\end{equation}
The different field amplitudes are plotted in Fig.~\ref{fig:beta1}(a) as a function of $\beta$ for the emitter located at an anti-node of the resonator ($\alpha_0=\pi$). For small emitter-mode channeling efficiencies $\beta\ll1$, the predictions for the four cavity fields $|\phi_1|...|\phi_4|$ from our cascaded model and the prediction of the global cavity field $|\phi_{\textrm{JC}}|$ from the JC model agree with each other. With increasing $\beta$, only the resonator fields on the right of the emitter, $|\phi_2|$ and $|\phi_3|$, approximately follow the predictions of the JC model while the fields $|\phi_1|$ and $|\phi_4|$ exhibit orders of magnitude different values. A striking difference between the predictions of the two models occurs for $\beta=1/3$ ($ \gamma_{l} = 2/3 \gamma $) where the emitter and the right mirror form
a critically coupled cavity, that scatters all the incident light into the environment. Consequently, $|\phi_4|$ drops to zero and no light couples back out of the cavity and the cavity reflection is simply given by the mirror reflection coefficient $r_1$. Furthermore, for $\beta\rightarrow 1$ ($ \gamma_{l} \rightarrow 0 $) the JC model predicts zero amplitude for the intracavity field, while in our cascaded model only $|\phi_2|$ and $|\phi_3|$ approach 0 while $|\phi_1|$ and $|\phi_4|$ approach $t_1/4$, the value predicted for an empty, one-sided resonator, that is maximally detuned from the probe laser. In this situation, the emitter acts as a perfect mirror that forms, together with the incoupling mirror, a resonator which prevents any light from reaching the second part of the cavity. 
Interestingly, the predictions of both models for the emitter's excitation amplitude, 
\begin{equation}
\phi_0/\phi_{in}=-i\frac{t_1}{2 \sin(\alpha_0/2)}\sqrt{\frac{c}{\beta\gamma}}\;,
\end{equation}
agree with each other over the whole $\beta$-range despite the large difference in the predicted photon flux in the cavity. This might also be a reason why this effect has been overlooked so far. Fig.~\ref{fig:beta1} (b) and (c) show the dependence of the cavity fields on the detuning between probe light ($\omega_p$) and the emitter which is resonant with the cavity ($\omega_0=\omega_a$) for the $\beta=1/3$ and $\beta=1$, respectively. As can be seen, the deviations between our cascaded model and the JC model are maximal at the resonance frequency of the emitter and reduce with increasing detuning. For large detunings, $|\omega_p-\omega_0|\gg\gamma$, the effect of the emitter on the cavity field is negligible for a single rountrip and the JC approximation and our cascaded model show very good agreement.

\begin{figure}[tb]
	\includegraphics[width=0.95\columnwidth]{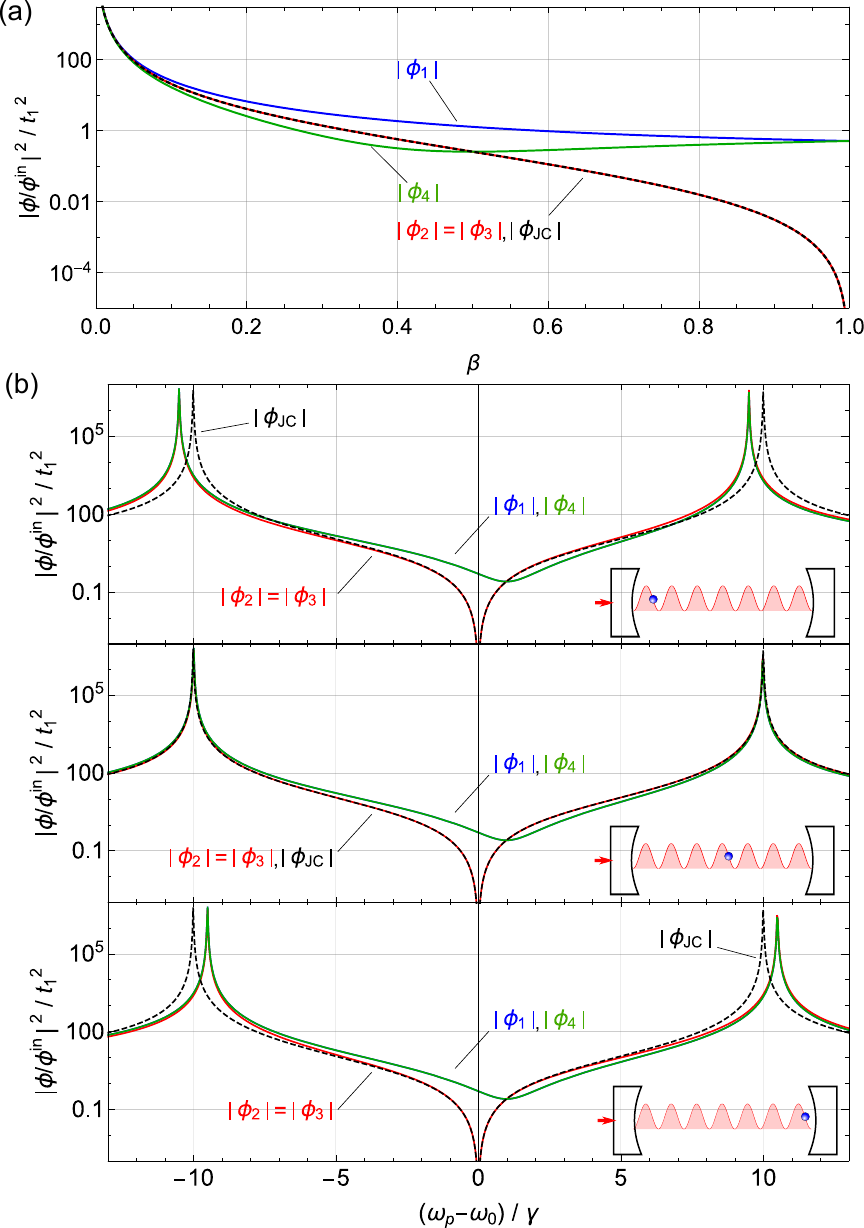}	
	\caption{(a) On-resonance field amplitude in the cavity as a function of emitter's channeling efficiency $\beta$ for a high-finesse resonator for an emitter positioned at a point of half-intensity of the cavity field mode ($\alpha_0=\pi/2$). The solid lines indicate the prediction of our cascaded model and the dashed line that of the JC model. (b)-(d) Excitation of the different cavity fields as a function of light-resonator detuning for an emitter positioned close to the incoupling mirror, the center of the cavity and the outcoupling mirror, respectively. For an emitter located at the other half-intensity position ({$ \alpha_0=-\pi/2 $}), we obtain a spectrum that is flipped with respect to $\omega_p-\omega_0=0$. The parameters of the calculation are $\beta=1$, $t_1^2=t_2^2=1-r_1^2=1-r_2^2=10^{-4}$ and $\nu_{\text{fsr} }=50\gamma$. }
	\label{fig:beta2}
\end{figure}

Now, we investigate the emitter-light interaction for the case where the emitter is not located at the intensity maximum of the standing wave resonator mode. Figure~\ref{fig:beta2}(a) shows the on-resonance excitation of the resonator fields according to our cascaded model and the Jaynes-Cummings predictions as a function of $\beta$ for an emitter located exactely between node and anti-node of the standing wave ($\alpha_0=\pi/2$), calculated from Eqs.~(\ref{eq:phi_res}). One observes similar discrepancies between the two models, however, the field no longer vanishes at $\beta=1/3$.

We now study the spectral response of the emitter-resonator system, and observe a dependence of the vacuum-Rabi spectrum on the emitter's position along the resonator ($x_a$) and with respect to the standing wave ($\alpha_0=\pm\pi/2$).
When the emitter is positioned on the rhs of the standing wave maximum ($\alpha_0=+\pi/2$), the cavity fields $ |\phi_1| $ and $ |\phi_4| $ are no longer minimal at the atomic resonance frequency, but one observes a frequency shift of their minimum position of $+\gamma\beta$ with respect to the atomic resonance, as shown in Figs.~\ref{fig:beta2}(b)-(d). Even more surprisingly, the spectral positions of the vacuum-Rabi resonances (maxima of $\phi_i $) of the coupled system are shifted by up to $\pm\Delta_R=\beta\gamma/2$ with respect to the prediction of the Jaynes-Cummings model depending on the axial position of the emitter $ x_a $ in the cavity. For an emitter located at the opposite flank of the standing wave ($\alpha_0=-\pi/2$), we obtain the spectra that are flipped with respect to $\omega_p-\omega_0=0$ compared to the ones shown in Figs.~\ref{fig:beta2}(b)-(d).
More generally, the frequency shift of the resonances with respect to the predictions of the JC model is given by
\begin{equation}
\Delta_R=\frac{\beta\gamma}{2}\sin\alpha_0\cdot (\frac{4x_a}{L}-1)=\frac{g^2_{max}}{8\nu_{\text{fsr} }}\sin\alpha_0\cdot (\frac{4x_a}{L}-1),
\end{equation}
see Appendix~\ref{chapter:alphadep}, where $g_{max}$ is the coupling strength at the cavity field anti-node ($\alpha_0=\pi$). Consequently, even though atom and cavity are in resonance, the vacuum-Rabi splitting is not symmetric with respect to the bare resonance. This asymmetry depends on the position of the emitter in the cavity's standing wave pattern ($x_a$ and $\alpha_0$) and $\beta$ and the shift of the resonances can reach s value of up to $\pm\gamma/2$. We note that, when probing the cavity through the other mirror, the spectral shifts will change sign which could offer a simple way to experimentally observe this effect.

The additional shift of the system's eigenfrequencies can be understood in a semi-classical picture. In this picture, the emitter-resonator coupling strength is governed by the intensity of standing wave mode at the emitter's position,  see $\alpha$ in Eq.~(\ref{eq:alpha}) in the Appendix. 
Thus, detuning the probe light from resonance, will result in a shifted standing wave pattern and, thus, depending on detuning and the emitter's position, in a decreased or increased coupling and a consequent shift of the Rabi peak.

In summary, we showed that in the situation of large channeling efficiency $\beta$ of a cavity-coupled emitter, the Jaynes-Cummings model does not correctly describe the physical situation and leads to quantitatively and qualitatively wrong predictions for settings that are already realized in many experiments. We present an alternative description based on a cascaded interaction of the propagating cavity field with the emitters that is valid for all parameter ranges of $\beta$. Interestingly, our results show that the parameter governing the deviations between the two models is not directly the emitter-resonator coupling strength $g$, but rather the channeling efficiency $\beta$ into the cavity mode. Therefore, significant deviations from the Jaynes-Cummings model can already occur for arbitrarily small values of $g$. We note that the cooperativity that is typically used to quantify the performance of a CQED system can be expressed in terms of $\beta$ as
\begin{equation}
C=\frac{g^2}{2\kappa_l\gamma_l}=\frac{2\beta}{1-\beta}\frac{F}{\pi}
\end{equation}
where $F=\pi\nu_{\text{fsr} }/\kappa_l$ is the finesse of the cavity. Thus, for given finesse, the only way to enhance the system's performance is to increase $\beta$, underpinning the practical relevance of our results. 

We note that in this manuscript, we limited our discussion to the weak driving regime. Further deviations from the JC model are expected when more than one excitation is present in the system. Then, it is not only the amplitude of the light field that will be position-dependent but also, e.g., the second or higher order correlation functions of the cavity field. The correct understanding of these effects may be crucial for improving our fundamental understanding of light-matter interaction and may open up paths towards novel applications. This is particularly important as many experiments strive for the high $ \beta$-regime where the field radiated by a single emitter has a large overlap with the cavity field mode.

We acknowledge financial support by the Austrian Science Fund (NanoFiRe Grant Project No. P 31115) and the Alexander von Humboldt Foundation in the framework of an Alexander von Humboldt Professorship endowed by the Federal Ministry of Education and Research.

\bibliography{JCpaper}

\begin{thebibliography}{38}%
\makeatletter
\providecommand \@ifxundefined [1]{%
 \@ifx{#1\undefined}
}%
\providecommand \@ifnum [1]{%
 \ifnum #1\expandafter \@firstoftwo
 \else \expandafter \@secondoftwo
 \fi
}%
\providecommand \@ifx [1]{%
 \ifx #1\expandafter \@firstoftwo
 \else \expandafter \@secondoftwo
 \fi
}%
\providecommand \natexlab [1]{#1}%
\providecommand \enquote  [1]{``#1''}%
\providecommand \bibnamefont  [1]{#1}%
\providecommand \bibfnamefont [1]{#1}%
\providecommand \citenamefont [1]{#1}%
\providecommand \href@noop [0]{\@secondoftwo}%
\providecommand \href [0]{\begingroup \@sanitize@url \@href}%
\providecommand \@href[1]{\@@startlink{#1}\@@href}%
\providecommand \@@href[1]{\endgroup#1\@@endlink}%
\providecommand \@sanitize@url [0]{\catcode `\\12\catcode `\$12\catcode
  `\&12\catcode `\#12\catcode `\^12\catcode `\_12\catcode `\%12\relax}%
\providecommand \@@startlink[1]{}%
\providecommand \@@endlink[0]{}%
\providecommand \url  [0]{\begingroup\@sanitize@url \@url }%
\providecommand \@url [1]{\endgroup\@href {#1}{\urlprefix }}%
\providecommand \urlprefix  [0]{URL }%
\providecommand \Eprint [0]{\href }%
\providecommand \doibase [0]{https://doi.org/}%
\providecommand \selectlanguage [0]{\@gobble}%
\providecommand \bibinfo  [0]{\@secondoftwo}%
\providecommand \bibfield  [0]{\@secondoftwo}%
\providecommand \translation [1]{[#1]}%
\providecommand \BibitemOpen [0]{}%
\providecommand \bibitemStop [0]{}%
\providecommand \bibitemNoStop [0]{.\EOS\space}%
\providecommand \EOS [0]{\spacefactor3000\relax}%
\providecommand \BibitemShut  [1]{\csname bibitem#1\endcsname}%
\let\auto@bib@innerbib\@empty
\bibitem [{\citenamefont {Jaynes}\ and\ \citenamefont
  {Cummings}(1963)}]{jaynes_comparison_1963}%
  \BibitemOpen
  \bibfield  {author} {\bibinfo {author} {\bibfnamefont {E.}~\bibnamefont
  {Jaynes}}\ and\ \bibinfo {author} {\bibfnamefont {F.}~\bibnamefont
  {Cummings}},\ }\bibfield  {title} {\bibinfo {title} {Comparison of quantum
  and semiclassical radiation theories with application to the beam maser},\
  }\href {https://doi.org/10.1109/PROC.1963.1664} {\bibfield  {journal}
  {\bibinfo  {journal} {Proceedings of the IEEE}\ }\textbf {\bibinfo {volume}
  {51}},\ \bibinfo {pages} {89} (\bibinfo {year} {1963})}\BibitemShut {NoStop}%
\bibitem [{\citenamefont {Walther}\ \emph {et~al.}(2006)\citenamefont
  {Walther}, \citenamefont {Varcoe}, \citenamefont {Englert},\ and\
  \citenamefont {Becker}}]{walther2006}%
  \BibitemOpen
  \bibfield  {author} {\bibinfo {author} {\bibfnamefont {H.}~\bibnamefont
  {Walther}}, \bibinfo {author} {\bibfnamefont {B.~T.~H.}\ \bibnamefont
  {Varcoe}}, \bibinfo {author} {\bibfnamefont {B.-G.}\ \bibnamefont
  {Englert}},\ and\ \bibinfo {author} {\bibfnamefont {T.}~\bibnamefont
  {Becker}},\ }\bibfield  {title} {\bibinfo {title} {Cavity quantum
  electrodynamics},\ }\href {https://doi.org/10.1088/0034-4885/69/5/r02}
  {\bibfield  {journal} {\bibinfo  {journal} {Rep. Prog. Phys.}\ }\textbf
  {\bibinfo {volume} {69}},\ \bibinfo {pages} {1325} (\bibinfo {year}
  {2006})}\BibitemShut {NoStop}%
\bibitem [{\citenamefont {Ye}\ \emph {et~al.}(1999)\citenamefont {Ye},
  \citenamefont {Vernooy},\ and\ \citenamefont {Kimble}}]{Ye1999}%
  \BibitemOpen
  \bibfield  {author} {\bibinfo {author} {\bibfnamefont {J.}~\bibnamefont
  {Ye}}, \bibinfo {author} {\bibfnamefont {D.~W.}\ \bibnamefont {Vernooy}},\
  and\ \bibinfo {author} {\bibfnamefont {H.~J.}\ \bibnamefont {Kimble}},\
  }\bibfield  {title} {\bibinfo {title} {Trapping of single atoms in cavity
  qed},\ }\href {https://doi.org/10.1103/PhysRevLett.83.4987} {\bibfield
  {journal} {\bibinfo  {journal} {Phys. Rev. Lett.}\ }\textbf {\bibinfo
  {volume} {83}},\ \bibinfo {pages} {4987} (\bibinfo {year}
  {1999})}\BibitemShut {NoStop}%
\bibitem [{\citenamefont {Pinkse}\ \emph {et~al.}(2000)\citenamefont {Pinkse},
  \citenamefont {Fischer}, \citenamefont {Maunz},\ and\ \citenamefont
  {Rempe}}]{Pinkse2000}%
  \BibitemOpen
  \bibfield  {author} {\bibinfo {author} {\bibfnamefont {P.~W.~H.}\
  \bibnamefont {Pinkse}}, \bibinfo {author} {\bibfnamefont {T.}~\bibnamefont
  {Fischer}}, \bibinfo {author} {\bibfnamefont {P.}~\bibnamefont {Maunz}},\
  and\ \bibinfo {author} {\bibfnamefont {G.}~\bibnamefont {Rempe}},\ }\bibfield
   {title} {\bibinfo {title} {Trapping an atom with single photons},\ }\href
  {https://doi.org/https://doi.org/10.1038/35006006} {\bibfield  {journal}
  {\bibinfo  {journal} {Nature}\ }\textbf {\bibinfo {volume} {404}},\ \bibinfo
  {pages} {365} (\bibinfo {year} {2000})}\BibitemShut {NoStop}%
\bibitem [{\citenamefont {Brune}\ \emph {et~al.}(1996)\citenamefont {Brune},
  \citenamefont {Schmidt-Kaler}, \citenamefont {Maali}, \citenamefont {Dreyer},
  \citenamefont {Hagley}, \citenamefont {Raimond},\ and\ \citenamefont
  {Haroche}}]{brune_quantum_1996}%
  \BibitemOpen
  \bibfield  {author} {\bibinfo {author} {\bibfnamefont {M.}~\bibnamefont
  {Brune}}, \bibinfo {author} {\bibfnamefont {F.}~\bibnamefont
  {Schmidt-Kaler}}, \bibinfo {author} {\bibfnamefont {A.}~\bibnamefont
  {Maali}}, \bibinfo {author} {\bibfnamefont {J.}~\bibnamefont {Dreyer}},
  \bibinfo {author} {\bibfnamefont {E.}~\bibnamefont {Hagley}}, \bibinfo
  {author} {\bibfnamefont {J.~M.}\ \bibnamefont {Raimond}},\ and\ \bibinfo
  {author} {\bibfnamefont {S.}~\bibnamefont {Haroche}},\ }\bibfield  {title}
  {\bibinfo {title} {{Quantum Rabi Oscillation: A Direct Test of Field
  Quantization in a Cavity}},\ }\href
  {https://doi.org/10.1103/PhysRevLett.76.1800} {\bibfield  {journal} {\bibinfo
   {journal} {Physical Review Letters}\ }\textbf {\bibinfo {volume} {76}},\
  \bibinfo {pages} {1800} (\bibinfo {year} {1996})}\BibitemShut {NoStop}%
\bibitem [{\citenamefont {Aoki}\ \emph {et~al.}(2006)\citenamefont {Aoki},
  \citenamefont {Dayan}, \citenamefont {Wilcut}, \citenamefont {Bowen},
  \citenamefont {Parkins}, \citenamefont {Kippenberg}, \citenamefont {Vahala},\
  and\ \citenamefont {Kimble}}]{Aoki2006}%
  \BibitemOpen
  \bibfield  {author} {\bibinfo {author} {\bibfnamefont {T.}~\bibnamefont
  {Aoki}}, \bibinfo {author} {\bibfnamefont {B.}~\bibnamefont {Dayan}},
  \bibinfo {author} {\bibfnamefont {E.}~\bibnamefont {Wilcut}}, \bibinfo
  {author} {\bibfnamefont {W.~P.}\ \bibnamefont {Bowen}}, \bibinfo {author}
  {\bibfnamefont {A.~S.}\ \bibnamefont {Parkins}}, \bibinfo {author}
  {\bibfnamefont {T.~J.}\ \bibnamefont {Kippenberg}}, \bibinfo {author}
  {\bibfnamefont {K.~J.}\ \bibnamefont {Vahala}},\ and\ \bibinfo {author}
  {\bibfnamefont {H.~J.}\ \bibnamefont {Kimble}},\ }\bibfield  {title}
  {\bibinfo {title} {{Observation of strong coupling between one atom and a
  monolithic microresonator}},\ }\href {https://doi.org/10.1038/nature05147}
  {\bibfield  {journal} {\bibinfo  {journal} {Nature}\ }\textbf {\bibinfo
  {volume} {443}},\ \bibinfo {pages} {671} (\bibinfo {year} {2006})},\ \Eprint
  {https://arxiv.org/abs/0606033} {arXiv:0606033 [quant-ph]} \BibitemShut
  {NoStop}%
\bibitem [{\citenamefont {Thompson}\ \emph {et~al.}(2013)\citenamefont
  {Thompson}, \citenamefont {Tiecke}, \citenamefont {{De Leon}}, \citenamefont
  {Feist}, \citenamefont {Akimov}, \citenamefont {Gullans}, \citenamefont
  {Zibrov}, \citenamefont {Vuleti{\'{c}}},\ and\ \citenamefont
  {Lukin}}]{Thompson2013}%
  \BibitemOpen
  \bibfield  {author} {\bibinfo {author} {\bibfnamefont {J.~D.}\ \bibnamefont
  {Thompson}}, \bibinfo {author} {\bibfnamefont {T.~G.}\ \bibnamefont
  {Tiecke}}, \bibinfo {author} {\bibfnamefont {N.~P.}\ \bibnamefont {{De
  Leon}}}, \bibinfo {author} {\bibfnamefont {J.}~\bibnamefont {Feist}},
  \bibinfo {author} {\bibfnamefont {A.~V.}\ \bibnamefont {Akimov}}, \bibinfo
  {author} {\bibfnamefont {M.}~\bibnamefont {Gullans}}, \bibinfo {author}
  {\bibfnamefont {A.~S.}\ \bibnamefont {Zibrov}}, \bibinfo {author}
  {\bibfnamefont {V.}~\bibnamefont {Vuleti{\'{c}}}},\ and\ \bibinfo {author}
  {\bibfnamefont {M.~D.}\ \bibnamefont {Lukin}},\ }\bibfield  {title} {\bibinfo
  {title} {{Coupling a single trapped atom to a nanoscale optical cavity}},\
  }\href {https://doi.org/10.1126/science.1237125} {\bibfield  {journal}
  {\bibinfo  {journal} {Science}\ }\textbf {\bibinfo {volume} {340}},\ \bibinfo
  {pages} {1202} (\bibinfo {year} {2013})}\BibitemShut {NoStop}%
\bibitem [{\citenamefont {Junge}\ \emph {et~al.}(2013)\citenamefont {Junge},
  \citenamefont {O'Shea}, \citenamefont {Volz},\ and\ \citenamefont
  {Rauschenbeutel}}]{junge2013}%
  \BibitemOpen
  \bibfield  {author} {\bibinfo {author} {\bibfnamefont {C.}~\bibnamefont
  {Junge}}, \bibinfo {author} {\bibfnamefont {D.}~\bibnamefont {O'Shea}},
  \bibinfo {author} {\bibfnamefont {J.}~\bibnamefont {Volz}},\ and\ \bibinfo
  {author} {\bibfnamefont {A.}~\bibnamefont {Rauschenbeutel}},\ }\bibfield
  {title} {\bibinfo {title} {Strong coupling between single atoms and
  nontransversal photons},\ }\href
  {https://doi.org/10.1103/PhysRevLett.110.213604} {\bibfield  {journal}
  {\bibinfo  {journal} {Phys. Rev. Lett.}\ }\textbf {\bibinfo {volume} {110}},\
  \bibinfo {pages} {213604} (\bibinfo {year} {2013})}\BibitemShut {NoStop}%
\bibitem [{\citenamefont {Santhosh}\ \emph {et~al.}(2016)\citenamefont
  {Santhosh}, \citenamefont {Bitton}, \citenamefont {Chuntonov},\ and\
  \citenamefont {Haran}}]{santhosh2016vacuum}%
  \BibitemOpen
  \bibfield  {author} {\bibinfo {author} {\bibfnamefont {K.}~\bibnamefont
  {Santhosh}}, \bibinfo {author} {\bibfnamefont {O.}~\bibnamefont {Bitton}},
  \bibinfo {author} {\bibfnamefont {L.}~\bibnamefont {Chuntonov}},\ and\
  \bibinfo {author} {\bibfnamefont {G.}~\bibnamefont {Haran}},\ }\bibfield
  {title} {\bibinfo {title} {Vacuum rabi splitting in a plasmonic cavity at the
  single quantum emitter limit},\ }\href
  {https://doi.org/https://doi.org/10.1038/ncomms11823} {\bibfield  {journal}
  {\bibinfo  {journal} {Nature communications}\ }\textbf {\bibinfo {volume}
  {7}},\ \bibinfo {pages} {1} (\bibinfo {year} {2016})}\BibitemShut {NoStop}%
\bibitem [{\citenamefont {Hugall}\ \emph {et~al.}(2018)\citenamefont {Hugall},
  \citenamefont {Singh},\ and\ \citenamefont {van
  Hulst}}]{hugall2018plasmonic}%
  \BibitemOpen
  \bibfield  {author} {\bibinfo {author} {\bibfnamefont {J.~T.}\ \bibnamefont
  {Hugall}}, \bibinfo {author} {\bibfnamefont {A.}~\bibnamefont {Singh}},\ and\
  \bibinfo {author} {\bibfnamefont {N.~F.}\ \bibnamefont {van Hulst}},\
  }\bibfield  {title} {\bibinfo {title} {Plasmonic cavity coupling},\ }\href
  {https://doi.org/https://doi.org/10.1021/acsphotonics.7b01139} {\bibfield
  {journal} {\bibinfo  {journal} {Acs Photonics}\ }\textbf {\bibinfo {volume}
  {5}},\ \bibinfo {pages} {43} (\bibinfo {year} {2018})}\BibitemShut {NoStop}%
\bibitem [{\citenamefont {Bitton}\ \emph {et~al.}(2020)\citenamefont {Bitton},
  \citenamefont {Gupta}, \citenamefont {Houben}, \citenamefont {Kvapil},
  \citenamefont {K{\v{r}}{\'a}pek}, \citenamefont {{\v{S}}ikola},\ and\
  \citenamefont {Haran}}]{bitton2020vacuum}%
  \BibitemOpen
  \bibfield  {author} {\bibinfo {author} {\bibfnamefont {O.}~\bibnamefont
  {Bitton}}, \bibinfo {author} {\bibfnamefont {S.~N.}\ \bibnamefont {Gupta}},
  \bibinfo {author} {\bibfnamefont {L.}~\bibnamefont {Houben}}, \bibinfo
  {author} {\bibfnamefont {M.}~\bibnamefont {Kvapil}}, \bibinfo {author}
  {\bibfnamefont {V.}~\bibnamefont {K{\v{r}}{\'a}pek}}, \bibinfo {author}
  {\bibfnamefont {T.}~\bibnamefont {{\v{S}}ikola}},\ and\ \bibinfo {author}
  {\bibfnamefont {G.}~\bibnamefont {Haran}},\ }\bibfield  {title} {\bibinfo
  {title} {Vacuum rabi splitting of a dark plasmonic cavity mode revealed by
  fast electrons},\ }\href
  {https://doi.org/https://doi.org/10.1038/s41467-020-14364-3} {\bibfield
  {journal} {\bibinfo  {journal} {Nature communications}\ }\textbf {\bibinfo
  {volume} {11}},\ \bibinfo {pages} {1} (\bibinfo {year} {2020})}\BibitemShut
  {NoStop}%
\bibitem [{\citenamefont {Schoelkopf}\ and\ \citenamefont
  {Girvin}(2008)}]{schoelkopf2008wiring}%
  \BibitemOpen
  \bibfield  {author} {\bibinfo {author} {\bibfnamefont {R.}~\bibnamefont
  {Schoelkopf}}\ and\ \bibinfo {author} {\bibfnamefont {S.}~\bibnamefont
  {Girvin}},\ }\bibfield  {title} {\bibinfo {title} {Wiring up quantum
  systems},\ }\href {https://doi.org/https://doi.org/10.1038/451664a}
  {\bibfield  {journal} {\bibinfo  {journal} {Nature}\ }\textbf {\bibinfo
  {volume} {451}},\ \bibinfo {pages} {664} (\bibinfo {year}
  {2008})}\BibitemShut {NoStop}%
\bibitem [{\citenamefont {Hu}\ \emph {et~al.}(2012)\citenamefont {Hu},
  \citenamefont {Liu},\ and\ \citenamefont {Nori}}]{Hu2012}%
  \BibitemOpen
  \bibfield  {author} {\bibinfo {author} {\bibfnamefont {X.}~\bibnamefont
  {Hu}}, \bibinfo {author} {\bibfnamefont {Y.-x.}\ \bibnamefont {Liu}},\ and\
  \bibinfo {author} {\bibfnamefont {F.}~\bibnamefont {Nori}},\ }\bibfield
  {title} {\bibinfo {title} {Strong coupling of a spin qubit to a
  superconducting stripline cavity},\ }\href
  {https://doi.org/10.1103/PhysRevB.86.035314} {\bibfield  {journal} {\bibinfo
  {journal} {Phys. Rev. B}\ }\textbf {\bibinfo {volume} {86}},\ \bibinfo
  {pages} {035314} (\bibinfo {year} {2012})}\BibitemShut {NoStop}%
\bibitem [{\citenamefont {Wang}\ \emph {et~al.}(2019)\citenamefont {Wang},
  \citenamefont {Kelkar}, \citenamefont {Martin-Cano}, \citenamefont
  {Rattenbacher}, \citenamefont {Shkarin}, \citenamefont {Utikal},
  \citenamefont {Götzinger},\ and\ \citenamefont {Sandoghdar}}]{wang_2019}%
  \BibitemOpen
  \bibfield  {author} {\bibinfo {author} {\bibfnamefont {D.}~\bibnamefont
  {Wang}}, \bibinfo {author} {\bibfnamefont {H.}~\bibnamefont {Kelkar}},
  \bibinfo {author} {\bibfnamefont {D.}~\bibnamefont {Martin-Cano}}, \bibinfo
  {author} {\bibfnamefont {D.}~\bibnamefont {Rattenbacher}}, \bibinfo {author}
  {\bibfnamefont {A.}~\bibnamefont {Shkarin}}, \bibinfo {author} {\bibfnamefont
  {T.}~\bibnamefont {Utikal}}, \bibinfo {author} {\bibfnamefont
  {S.}~\bibnamefont {Götzinger}},\ and\ \bibinfo {author} {\bibfnamefont
  {V.}~\bibnamefont {Sandoghdar}},\ }\bibfield  {title} {\bibinfo {title}
  {Turning a molecule into a coherent two-level quantum system},\ }\href
  {https://doi.org/10.1038/s41567-019-0436-5} {\bibfield  {journal} {\bibinfo
  {journal} {Nature Physics}\ }\textbf {\bibinfo {volume} {15}},\ \bibinfo
  {pages} {483} (\bibinfo {year} {2019})}\BibitemShut {NoStop}%
\bibitem [{\citenamefont {Michler}\ \emph {et~al.}(2000)\citenamefont
  {Michler}, \citenamefont {Kiraz}, \citenamefont {Becher}, \citenamefont
  {Schoenfeld}, \citenamefont {Petroff}, \citenamefont {Zhang}, \citenamefont
  {Hu},\ and\ \citenamefont {Imamoglu}}]{michler2000quantum}%
  \BibitemOpen
  \bibfield  {author} {\bibinfo {author} {\bibfnamefont {P.}~\bibnamefont
  {Michler}}, \bibinfo {author} {\bibfnamefont {A.}~\bibnamefont {Kiraz}},
  \bibinfo {author} {\bibfnamefont {C.}~\bibnamefont {Becher}}, \bibinfo
  {author} {\bibfnamefont {W.}~\bibnamefont {Schoenfeld}}, \bibinfo {author}
  {\bibfnamefont {P.}~\bibnamefont {Petroff}}, \bibinfo {author} {\bibfnamefont
  {L.}~\bibnamefont {Zhang}}, \bibinfo {author} {\bibfnamefont
  {E.}~\bibnamefont {Hu}},\ and\ \bibinfo {author} {\bibfnamefont
  {A.}~\bibnamefont {Imamoglu}},\ }\bibfield  {title} {\bibinfo {title} {A
  quantum dot single-photon turnstile device},\ }\href
  {https://doi.org/10.1126/science.290.5500.2282} {\bibfield  {journal}
  {\bibinfo  {journal} {Science}\ }\textbf {\bibinfo {volume} {290}},\ \bibinfo
  {pages} {2282} (\bibinfo {year} {2000})}\BibitemShut {NoStop}%
\bibitem [{\citenamefont {Yoshie}\ \emph {et~al.}(2004)\citenamefont {Yoshie},
  \citenamefont {Scherer}, \citenamefont {Hendrickson}, \citenamefont
  {Khitrova}, \citenamefont {Gibbs}, \citenamefont {Rupper}, \citenamefont
  {Ell}, \citenamefont {Shchekin},\ and\ \citenamefont
  {Deppe}}]{yoshie_vacuum_2004}%
  \BibitemOpen
  \bibfield  {author} {\bibinfo {author} {\bibfnamefont {T.}~\bibnamefont
  {Yoshie}}, \bibinfo {author} {\bibfnamefont {A.}~\bibnamefont {Scherer}},
  \bibinfo {author} {\bibfnamefont {J.}~\bibnamefont {Hendrickson}}, \bibinfo
  {author} {\bibfnamefont {G.}~\bibnamefont {Khitrova}}, \bibinfo {author}
  {\bibfnamefont {H.~M.}\ \bibnamefont {Gibbs}}, \bibinfo {author}
  {\bibfnamefont {G.}~\bibnamefont {Rupper}}, \bibinfo {author} {\bibfnamefont
  {C.}~\bibnamefont {Ell}}, \bibinfo {author} {\bibfnamefont {O.~B.}\
  \bibnamefont {Shchekin}},\ and\ \bibinfo {author} {\bibfnamefont {D.~G.}\
  \bibnamefont {Deppe}},\ }\bibfield  {title} {\bibinfo {title} {Vacuum {Rabi}
  splitting with a single quantum dot in a photonic crystal nanocavity},\
  }\href {https://doi.org/10.1038/nature03119} {\bibfield  {journal} {\bibinfo
  {journal} {Nature}\ }\textbf {\bibinfo {volume} {432}},\ \bibinfo {pages}
  {200} (\bibinfo {year} {2004})}\BibitemShut {NoStop}%
\bibitem [{\citenamefont {Merkel}\ \emph {et~al.}(2020)\citenamefont {Merkel},
  \citenamefont {Ulanowski},\ and\ \citenamefont {Reiserer}}]{Merkel2020}%
  \BibitemOpen
  \bibfield  {author} {\bibinfo {author} {\bibfnamefont {B.}~\bibnamefont
  {Merkel}}, \bibinfo {author} {\bibfnamefont {A.}~\bibnamefont {Ulanowski}},\
  and\ \bibinfo {author} {\bibfnamefont {A.}~\bibnamefont {Reiserer}},\
  }\bibfield  {title} {\bibinfo {title} {{Coherent and Purcell-Enhanced
  Emission from Erbium Dopants in a Cryogenic High- Q Resonator}},\ }\href
  {https://doi.org/10.1103/PhysRevX.10.041025} {\bibfield  {journal} {\bibinfo
  {journal} {Phys. Rev. X}\ }\textbf {\bibinfo {volume} {10}},\ \bibinfo
  {pages} {41025} (\bibinfo {year} {2020})},\ \Eprint
  {https://arxiv.org/abs/2006.14229} {arXiv:2006.14229} \BibitemShut {NoStop}%
\bibitem [{\citenamefont {Zhang}\ \emph {et~al.}(2018)\citenamefont {Zhang},
  \citenamefont {Sun}, \citenamefont {Burek}, \citenamefont {Dory},
  \citenamefont {Tzeng}, \citenamefont {Fischer}, \citenamefont {Kelaita},
  \citenamefont {Lagoudakis}, \citenamefont {Radulaski}, \citenamefont {Shen},
  \citenamefont {Meloshi}, \citenamefont {Chu}, \citenamefont {Loncar},\ and\
  \citenamefont {Vukovic}}]{zhang2018strongly}%
  \BibitemOpen
  \bibfield  {author} {\bibinfo {author} {\bibfnamefont {J.~L.}\ \bibnamefont
  {Zhang}}, \bibinfo {author} {\bibfnamefont {S.}~\bibnamefont {Sun}}, \bibinfo
  {author} {\bibfnamefont {M.~J.}\ \bibnamefont {Burek}}, \bibinfo {author}
  {\bibfnamefont {C.}~\bibnamefont {Dory}}, \bibinfo {author} {\bibfnamefont
  {Y.-K.}\ \bibnamefont {Tzeng}}, \bibinfo {author} {\bibfnamefont {K.~A.}\
  \bibnamefont {Fischer}}, \bibinfo {author} {\bibfnamefont {Y.}~\bibnamefont
  {Kelaita}}, \bibinfo {author} {\bibfnamefont {K.~G.}\ \bibnamefont
  {Lagoudakis}}, \bibinfo {author} {\bibfnamefont {M.}~\bibnamefont
  {Radulaski}}, \bibinfo {author} {\bibfnamefont {Z.-X.}\ \bibnamefont {Shen}},
  \bibinfo {author} {\bibfnamefont {N.}~\bibnamefont {Meloshi}}, \bibinfo
  {author} {\bibfnamefont {S.}~\bibnamefont {Chu}}, \bibinfo {author}
  {\bibfnamefont {M.}~\bibnamefont {Loncar}},\ and\ \bibinfo {author}
  {\bibfnamefont {J.}~\bibnamefont {Vukovic}},\ }\bibfield  {title} {\bibinfo
  {title} {Strongly cavity-enhanced spontaneous emission from silicon-vacancy
  centers in diamond},\ }\href
  {https://doi.org/https://doi.org/10.1021/acs.nanolett.7b05075} {\bibfield
  {journal} {\bibinfo  {journal} {Nano letters}\ }\textbf {\bibinfo {volume}
  {18}},\ \bibinfo {pages} {1360} (\bibinfo {year} {2018})}\BibitemShut
  {NoStop}%
\bibitem [{\citenamefont {Fink}\ \emph {et~al.}(2008)\citenamefont {Fink},
  \citenamefont {Göppl}, \citenamefont {Baur}, \citenamefont {Bianchetti},
  \citenamefont {Leek}, \citenamefont {Blais},\ and\ \citenamefont
  {Wallraff}}]{fink_2008}%
  \BibitemOpen
  \bibfield  {author} {\bibinfo {author} {\bibfnamefont {J.~M.}\ \bibnamefont
  {Fink}}, \bibinfo {author} {\bibfnamefont {M.}~\bibnamefont {Göppl}},
  \bibinfo {author} {\bibfnamefont {M.}~\bibnamefont {Baur}}, \bibinfo {author}
  {\bibfnamefont {R.}~\bibnamefont {Bianchetti}}, \bibinfo {author}
  {\bibfnamefont {P.~J.}\ \bibnamefont {Leek}}, \bibinfo {author}
  {\bibfnamefont {A.}~\bibnamefont {Blais}},\ and\ \bibinfo {author}
  {\bibfnamefont {A.}~\bibnamefont {Wallraff}},\ }\bibfield  {title} {\bibinfo
  {title} {Climbing the {Jaynes}–{Cummings} ladder and observing its
  nonlinearity in a cavity {QED} system},\ }\href
  {https://doi.org/10.1038/nature07112} {\bibfield  {journal} {\bibinfo
  {journal} {Nature}\ }\textbf {\bibinfo {volume} {454}},\ \bibinfo {pages}
  {315} (\bibinfo {year} {2008})}\BibitemShut {NoStop}%
\bibitem [{\citenamefont {Reiserer}\ and\ \citenamefont
  {Rempe}(2015)}]{reiserer_cavity-based_2015}%
  \BibitemOpen
  \bibfield  {author} {\bibinfo {author} {\bibfnamefont {A.}~\bibnamefont
  {Reiserer}}\ and\ \bibinfo {author} {\bibfnamefont {G.}~\bibnamefont
  {Rempe}},\ }\bibfield  {title} {\bibinfo {title} {Cavity-based quantum
  networks with single atoms and optical photons},\ }\href
  {https://doi.org/10.1103/RevModPhys.87.1379} {\bibfield  {journal} {\bibinfo
  {journal} {Reviews of Modern Physics}\ }\textbf {\bibinfo {volume} {87}},\
  \bibinfo {pages} {1379} (\bibinfo {year} {2015})}\BibitemShut {NoStop}%
\bibitem [{\citenamefont {Meiser}\ and\ \citenamefont
  {Meystre}(2006)}]{meiser_superstrong_2006}%
  \BibitemOpen
  \bibfield  {author} {\bibinfo {author} {\bibfnamefont {D.}~\bibnamefont
  {Meiser}}\ and\ \bibinfo {author} {\bibfnamefont {P.}~\bibnamefont
  {Meystre}},\ }\bibfield  {title} {\bibinfo {title} {{Superstrong coupling
  regime of cavity quantum electrodynamics}},\ }\href
  {https://doi.org/10.1103/PhysRevA.74.065801} {\bibfield  {journal} {\bibinfo
  {journal} {Phys. Rev. A}\ }\textbf {\bibinfo {volume} {74}},\ \bibinfo
  {pages} {6} (\bibinfo {year} {2006})}\BibitemShut {NoStop}%
\bibitem [{\citenamefont {Sundaresan}\ \emph {et~al.}(2015)\citenamefont
  {Sundaresan}, \citenamefont {Liu}, \citenamefont {Sadri}, \citenamefont
  {Sz{\"{o}}cs}, \citenamefont {Underwood}, \citenamefont {Malekakhlagh},
  \citenamefont {T{\"{u}}reci},\ and\ \citenamefont {Houck}}]{Sundaresan2015}%
  \BibitemOpen
  \bibfield  {author} {\bibinfo {author} {\bibfnamefont {N.~M.}\ \bibnamefont
  {Sundaresan}}, \bibinfo {author} {\bibfnamefont {Y.}~\bibnamefont {Liu}},
  \bibinfo {author} {\bibfnamefont {D.}~\bibnamefont {Sadri}}, \bibinfo
  {author} {\bibfnamefont {L.~J.}\ \bibnamefont {Sz{\"{o}}cs}}, \bibinfo
  {author} {\bibfnamefont {D.~L.}\ \bibnamefont {Underwood}}, \bibinfo {author}
  {\bibfnamefont {M.}~\bibnamefont {Malekakhlagh}}, \bibinfo {author}
  {\bibfnamefont {H.~E.}\ \bibnamefont {T{\"{u}}reci}},\ and\ \bibinfo {author}
  {\bibfnamefont {A.~A.}\ \bibnamefont {Houck}},\ }\bibfield  {title} {\bibinfo
  {title} {{Beyond Strong Coupling in a Multimode Cavity}},\ }\href
  {https://link.aps.org/doi/10.1103/PhysRevX.5.021035} {\bibfield  {journal}
  {\bibinfo  {journal} {Phys. Rev. X}\ }\textbf {\bibinfo {volume} {5}},\
  \bibinfo {pages} {21035} (\bibinfo {year} {2015})}\BibitemShut {NoStop}%
\bibitem [{\citenamefont {Yoshihara}\ \emph {et~al.}(2017)\citenamefont
  {Yoshihara}, \citenamefont {Fuse}, \citenamefont {Ashhab}, \citenamefont
  {Kakuyanagi}, \citenamefont {Saito},\ and\ \citenamefont
  {Semba}}]{yoshihara_superconducting_2017}%
  \BibitemOpen
  \bibfield  {author} {\bibinfo {author} {\bibfnamefont {F.}~\bibnamefont
  {Yoshihara}}, \bibinfo {author} {\bibfnamefont {T.}~\bibnamefont {Fuse}},
  \bibinfo {author} {\bibfnamefont {S.}~\bibnamefont {Ashhab}}, \bibinfo
  {author} {\bibfnamefont {K.}~\bibnamefont {Kakuyanagi}}, \bibinfo {author}
  {\bibfnamefont {S.}~\bibnamefont {Saito}},\ and\ \bibinfo {author}
  {\bibfnamefont {K.}~\bibnamefont {Semba}},\ }\bibfield  {title} {\bibinfo
  {title} {Superconducting qubit–oscillator circuit beyond the
  ultrastrong-coupling regime},\ }\href {https://doi.org/10.1038/nphys3906}
  {\bibfield  {journal} {\bibinfo  {journal} {Nat. Phys.}\ }\textbf {\bibinfo
  {volume} {13}},\ \bibinfo {pages} {44} (\bibinfo {year} {2017})}\BibitemShut
  {NoStop}%
\bibitem [{\citenamefont {Johnson}\ \emph {et~al.}(2019)\citenamefont
  {Johnson}, \citenamefont {Blaha}, \citenamefont {Ulanov}, \citenamefont
  {Rauschenbeutel}, \citenamefont {Schneeweiss},\ and\ \citenamefont
  {Volz}}]{johnson_observation_2019}%
  \BibitemOpen
  \bibfield  {author} {\bibinfo {author} {\bibfnamefont {A.}~\bibnamefont
  {Johnson}}, \bibinfo {author} {\bibfnamefont {M.}~\bibnamefont {Blaha}},
  \bibinfo {author} {\bibfnamefont {A.~E.}\ \bibnamefont {Ulanov}}, \bibinfo
  {author} {\bibfnamefont {A.}~\bibnamefont {Rauschenbeutel}}, \bibinfo
  {author} {\bibfnamefont {P.}~\bibnamefont {Schneeweiss}},\ and\ \bibinfo
  {author} {\bibfnamefont {J.}~\bibnamefont {Volz}},\ }\bibfield  {title}
  {\bibinfo {title} {Observation of {Collective} {Superstrong} {Coupling} of
  {Cold} {Atoms} to a 30-m {Long} {Optical} {Resonator}},\ }\href
  {https://doi.org/10.1103/PhysRevLett.123.243602} {\bibfield  {journal}
  {\bibinfo  {journal} {Physical Review Letters}\ }\textbf {\bibinfo {volume}
  {123}},\ \bibinfo {pages} {243602} (\bibinfo {year} {2019})}\BibitemShut
  {NoStop}%
\bibitem [{\citenamefont {Kockum}\ \emph {et~al.}(2019)\citenamefont {Kockum},
  \citenamefont {Miranowicz}, \citenamefont {De~Liberato}, \citenamefont
  {Savasta},\ and\ \citenamefont {Nori}}]{kockum2019}%
  \BibitemOpen
  \bibfield  {author} {\bibinfo {author} {\bibfnamefont {A.~F.}\ \bibnamefont
  {Kockum}}, \bibinfo {author} {\bibfnamefont {A.}~\bibnamefont {Miranowicz}},
  \bibinfo {author} {\bibfnamefont {S.}~\bibnamefont {De~Liberato}}, \bibinfo
  {author} {\bibfnamefont {S.}~\bibnamefont {Savasta}},\ and\ \bibinfo {author}
  {\bibfnamefont {F.}~\bibnamefont {Nori}},\ }\bibfield  {title} {\bibinfo
  {title} {Ultrastrong coupling between light and matter},\ }\href
  {https://doi.org/https://doi.org/10.1038/s42254-018-0006-2} {\bibfield
  {journal} {\bibinfo  {journal} {Nature Reviews Physics}\ }\textbf {\bibinfo
  {volume} {1}},\ \bibinfo {pages} {19} (\bibinfo {year} {2019})}\BibitemShut
  {NoStop}%
\bibitem [{\citenamefont {Forn-D\'{\i}az}\ \emph {et~al.}(2019)\citenamefont
  {Forn-D\'{\i}az}, \citenamefont {Lamata}, \citenamefont {Rico}, \citenamefont
  {Kono},\ and\ \citenamefont {Solano}}]{Ultrastrong_2019}%
  \BibitemOpen
  \bibfield  {author} {\bibinfo {author} {\bibfnamefont {P.}~\bibnamefont
  {Forn-D\'{\i}az}}, \bibinfo {author} {\bibfnamefont {L.}~\bibnamefont
  {Lamata}}, \bibinfo {author} {\bibfnamefont {E.}~\bibnamefont {Rico}},
  \bibinfo {author} {\bibfnamefont {J.}~\bibnamefont {Kono}},\ and\ \bibinfo
  {author} {\bibfnamefont {E.}~\bibnamefont {Solano}},\ }\bibfield  {title}
  {\bibinfo {title} {Ultrastrong coupling regimes of light-matter
  interaction},\ }\href {https://doi.org/10.1103/RevModPhys.91.025005}
  {\bibfield  {journal} {\bibinfo  {journal} {Rev. Mod. Phys.}\ }\textbf
  {\bibinfo {volume} {91}},\ \bibinfo {pages} {025005} (\bibinfo {year}
  {2019})}\BibitemShut {NoStop}%
\bibitem [{\citenamefont {Benedikter}\ \emph {et~al.}(2017)\citenamefont
  {Benedikter}, \citenamefont {Kaupp}, \citenamefont {H{\"{u}}mmer},
  \citenamefont {Liang}, \citenamefont {Bommer}, \citenamefont {Becher},
  \citenamefont {Krueger}, \citenamefont {Smith}, \citenamefont
  {H{\"{a}}nsch},\ and\ \citenamefont {Hunger}}]{Benedikter2017}%
  \BibitemOpen
  \bibfield  {author} {\bibinfo {author} {\bibfnamefont {J.}~\bibnamefont
  {Benedikter}}, \bibinfo {author} {\bibfnamefont {H.}~\bibnamefont {Kaupp}},
  \bibinfo {author} {\bibfnamefont {T.}~\bibnamefont {H{\"{u}}mmer}}, \bibinfo
  {author} {\bibfnamefont {Y.}~\bibnamefont {Liang}}, \bibinfo {author}
  {\bibfnamefont {A.}~\bibnamefont {Bommer}}, \bibinfo {author} {\bibfnamefont
  {C.}~\bibnamefont {Becher}}, \bibinfo {author} {\bibfnamefont
  {A.}~\bibnamefont {Krueger}}, \bibinfo {author} {\bibfnamefont {J.~M.}\
  \bibnamefont {Smith}}, \bibinfo {author} {\bibfnamefont {T.~W.}\ \bibnamefont
  {H{\"{a}}nsch}},\ and\ \bibinfo {author} {\bibfnamefont {D.}~\bibnamefont
  {Hunger}},\ }\bibfield  {title} {\bibinfo {title} {{Cavity-Enhanced
  Single-Photon Source Based on the Silicon-Vacancy Center in Diamond}},\
  }\href {https://doi.org/10.1103/PhysRevApplied.7.024031} {\bibfield
  {journal} {\bibinfo  {journal} {Phys. Rev. Appl.}\ }\textbf {\bibinfo
  {volume} {7}},\ \bibinfo {pages} {1} (\bibinfo {year} {2017})},\ \Eprint
  {https://arxiv.org/abs/1612.05509} {arXiv:1612.05509} \BibitemShut {NoStop}%
\bibitem [{\citenamefont {Gleyzes}\ \emph {et~al.}(2007)\citenamefont
  {Gleyzes}, \citenamefont {Kuhr}, \citenamefont {Guerlin}, \citenamefont
  {Bernu}, \citenamefont {Deleglise}, \citenamefont {Busk~Hoff}, \citenamefont
  {Brune}, \citenamefont {Raimond},\ and\ \citenamefont
  {Haroche}}]{gleyzes2007quantum}%
  \BibitemOpen
  \bibfield  {author} {\bibinfo {author} {\bibfnamefont {S.}~\bibnamefont
  {Gleyzes}}, \bibinfo {author} {\bibfnamefont {S.}~\bibnamefont {Kuhr}},
  \bibinfo {author} {\bibfnamefont {C.}~\bibnamefont {Guerlin}}, \bibinfo
  {author} {\bibfnamefont {J.}~\bibnamefont {Bernu}}, \bibinfo {author}
  {\bibfnamefont {S.}~\bibnamefont {Deleglise}}, \bibinfo {author}
  {\bibfnamefont {U.}~\bibnamefont {Busk~Hoff}}, \bibinfo {author}
  {\bibfnamefont {M.}~\bibnamefont {Brune}}, \bibinfo {author} {\bibfnamefont
  {J.-M.}\ \bibnamefont {Raimond}},\ and\ \bibinfo {author} {\bibfnamefont
  {S.}~\bibnamefont {Haroche}},\ }\bibfield  {title} {\bibinfo {title} {Quantum
  jumps of light recording the birth and death of a photon in a cavity},\
  }\href {https://doi.org/https://doi.org/10.1038/nature05589} {\bibfield
  {journal} {\bibinfo  {journal} {Nature}\ }\textbf {\bibinfo {volume} {446}},\
  \bibinfo {pages} {297} (\bibinfo {year} {2007})}\BibitemShut {NoStop}%
\bibitem [{\citenamefont {Le~Kien}\ and\ \citenamefont
  {Rauschenbeutel}(2016)}]{Fam_decay}%
  \BibitemOpen
  \bibfield  {author} {\bibinfo {author} {\bibfnamefont {F.}~\bibnamefont
  {Le~Kien}}\ and\ \bibinfo {author} {\bibfnamefont {A.}~\bibnamefont
  {Rauschenbeutel}},\ }\bibfield  {title} {\bibinfo {title} {Spontaneous
  emission of a two-level atom with an arbitrarily polarized electric dipole in
  front of a flat dielectric surface},\ }\href
  {https://doi.org/10.1103/PhysRevA.93.043828} {\bibfield  {journal} {\bibinfo
  {journal} {Phys. Rev. A}\ }\textbf {\bibinfo {volume} {93}},\ \bibinfo
  {pages} {043828} (\bibinfo {year} {2016})}\BibitemShut {NoStop}%
\bibitem [{\citenamefont {Skoff}\ \emph {et~al.}(2018)\citenamefont {Skoff},
  \citenamefont {Papencordt}, \citenamefont {Schauffert}, \citenamefont
  {Bayer},\ and\ \citenamefont {Rauschenbeutel}}]{Skoff2018}%
  \BibitemOpen
  \bibfield  {author} {\bibinfo {author} {\bibfnamefont {S.~M.}\ \bibnamefont
  {Skoff}}, \bibinfo {author} {\bibfnamefont {D.}~\bibnamefont {Papencordt}},
  \bibinfo {author} {\bibfnamefont {H.}~\bibnamefont {Schauffert}}, \bibinfo
  {author} {\bibfnamefont {B.~C.}\ \bibnamefont {Bayer}},\ and\ \bibinfo
  {author} {\bibfnamefont {A.}~\bibnamefont {Rauschenbeutel}},\ }\bibfield
  {title} {\bibinfo {title} {Optical-nanofiber-based interface for single
  molecules},\ }\href {https://doi.org/10.1103/PhysRevA.97.043839} {\bibfield
  {journal} {\bibinfo  {journal} {Phys. Rev. A}\ }\textbf {\bibinfo {volume}
  {97}},\ \bibinfo {pages} {043839} (\bibinfo {year} {2018})}\BibitemShut
  {NoStop}%
\bibitem [{\citenamefont {Bosman}\ \emph {et~al.}(2017)\citenamefont {Bosman},
  \citenamefont {Gely}, \citenamefont {Singh}, \citenamefont {Bruno},
  \citenamefont {Bothner},\ and\ \citenamefont {Steele}}]{bosman2017multi}%
  \BibitemOpen
  \bibfield  {author} {\bibinfo {author} {\bibfnamefont {S.~J.}\ \bibnamefont
  {Bosman}}, \bibinfo {author} {\bibfnamefont {M.~F.}\ \bibnamefont {Gely}},
  \bibinfo {author} {\bibfnamefont {V.}~\bibnamefont {Singh}}, \bibinfo
  {author} {\bibfnamefont {A.}~\bibnamefont {Bruno}}, \bibinfo {author}
  {\bibfnamefont {D.}~\bibnamefont {Bothner}},\ and\ \bibinfo {author}
  {\bibfnamefont {G.~A.}\ \bibnamefont {Steele}},\ }\bibfield  {title}
  {\bibinfo {title} {Multi-mode ultra-strong coupling in circuit quantum
  electrodynamics},\ }\href
  {https://doi.org/https://doi.org/10.1038/s41534-017-0046-y} {\bibfield
  {journal} {\bibinfo  {journal} {npj Quantum Information}\ }\textbf {\bibinfo
  {volume} {3}},\ \bibinfo {pages} {1} (\bibinfo {year} {2017})}\BibitemShut
  {NoStop}%
\bibitem [{\citenamefont {Kuzmin}\ \emph {et~al.}(2019)\citenamefont {Kuzmin},
  \citenamefont {Mehta}, \citenamefont {Grabon}, \citenamefont {Mencia},\ and\
  \citenamefont {Manucharyan}}]{Kuzmin2019}%
  \BibitemOpen
  \bibfield  {author} {\bibinfo {author} {\bibfnamefont {R.}~\bibnamefont
  {Kuzmin}}, \bibinfo {author} {\bibfnamefont {N.}~\bibnamefont {Mehta}},
  \bibinfo {author} {\bibfnamefont {N.}~\bibnamefont {Grabon}}, \bibinfo
  {author} {\bibfnamefont {R.}~\bibnamefont {Mencia}},\ and\ \bibinfo {author}
  {\bibfnamefont {V.~E.}\ \bibnamefont {Manucharyan}},\ }\bibfield  {title}
  {\bibinfo {title} {{Superstrong coupling in circuit quantum
  electrodynamics}},\ }\bibfield  {journal} {\bibinfo  {journal} {npj Quantum
  Inf.}\ }\textbf {\bibinfo {volume} {5}},\ \href
  {https://doi.org/10.1038/s41534-019-0134-2} {10.1038/s41534-019-0134-2}
  (\bibinfo {year} {2019}),\ \Eprint {https://arxiv.org/abs/1809.10739}
  {arXiv:1809.10739} \BibitemShut {NoStop}%
\bibitem [{\citenamefont {Shen}\ and\ \citenamefont {Fan}(2009)}]{Shen2009}%
  \BibitemOpen
  \bibfield  {author} {\bibinfo {author} {\bibfnamefont {J.-T.~T.}\
  \bibnamefont {Shen}}\ and\ \bibinfo {author} {\bibfnamefont {S.}~\bibnamefont
  {Fan}},\ }\bibfield  {title} {\bibinfo {title} {{Theory of single-photon
  transport in a single-mode waveguide. I. Coupling to a cavity containing a
  two-level atom}},\ }\href {https://doi.org/10.1103/PhysRevA.79.023837}
  {\bibfield  {journal} {\bibinfo  {journal} {Phys. Rev. A - At. Mol. Opt.
  Phys.}\ }\textbf {\bibinfo {volume} {79}},\ \bibinfo {pages} {1} (\bibinfo
  {year} {2009})}\BibitemShut {NoStop}%
\bibitem [{\citenamefont {Blaha}\ \emph {et~al.}(2022)\citenamefont {Blaha},
  \citenamefont {Johnson}, \citenamefont {Rauschenbeutel},\ and\ \citenamefont
  {Volz}}]{Blaha2021}%
  \BibitemOpen
  \bibfield  {author} {\bibinfo {author} {\bibfnamefont {M.}~\bibnamefont
  {Blaha}}, \bibinfo {author} {\bibfnamefont {A.}~\bibnamefont {Johnson}},
  \bibinfo {author} {\bibfnamefont {A.}~\bibnamefont {Rauschenbeutel}},\ and\
  \bibinfo {author} {\bibfnamefont {J.}~\bibnamefont {Volz}},\ }\bibfield
  {title} {\bibinfo {title} {{Beyond the Tavis-Cummings model: Revisiting
  cavity QED with ensembles of quantum emitters}},\ }\href
  {https://doi.org/10.1103/PhysRevA.105.013719} {\bibfield  {journal} {\bibinfo
   {journal} {Phys. Rev. A}\ }\textbf {\bibinfo {volume} {105}},\ \bibinfo
  {pages} {013719} (\bibinfo {year} {2022})}\BibitemShut {NoStop}%
\bibitem [{Note1()}]{Note1}%
  \BibitemOpen
  \bibinfo {note} {For fields propagating in a dispersionless medium the
  expression simplifies to $\partial _x=\partial /\partial x$}\BibitemShut
  {NoStop}%
\bibitem [{\citenamefont {Lodahl}\ \emph {et~al.}(2017)\citenamefont {Lodahl},
  \citenamefont {Mahmoodian}, \citenamefont {Stobbe}, \citenamefont
  {Schneeweiss}, \citenamefont {Volz}, \citenamefont {Rauschenbeutel},
  \citenamefont {Pichler}, \citenamefont {Zoller}, \citenamefont {Schneeweiss},
  \citenamefont {Volz}, \citenamefont {Pichler},\ and\ \citenamefont
  {Zoller}}]{lodahl_chiral_2017}%
  \BibitemOpen
  \bibfield  {author} {\bibinfo {author} {\bibfnamefont {P.}~\bibnamefont
  {Lodahl}}, \bibinfo {author} {\bibfnamefont {S.}~\bibnamefont {Mahmoodian}},
  \bibinfo {author} {\bibfnamefont {S.}~\bibnamefont {Stobbe}}, \bibinfo
  {author} {\bibfnamefont {P.}~\bibnamefont {Schneeweiss}}, \bibinfo {author}
  {\bibfnamefont {J.}~\bibnamefont {Volz}}, \bibinfo {author} {\bibfnamefont
  {A.}~\bibnamefont {Rauschenbeutel}}, \bibinfo {author} {\bibfnamefont
  {H.}~\bibnamefont {Pichler}}, \bibinfo {author} {\bibfnamefont
  {P.}~\bibnamefont {Zoller}}, \bibinfo {author} {\bibfnamefont
  {P.}~\bibnamefont {Schneeweiss}}, \bibinfo {author} {\bibfnamefont
  {J.}~\bibnamefont {Volz}}, \bibinfo {author} {\bibfnamefont {H.}~\bibnamefont
  {Pichler}},\ and\ \bibinfo {author} {\bibfnamefont {P.}~\bibnamefont
  {Zoller}},\ }\bibfield  {title} {\bibinfo {title} {{Chiral quantum optics}},\
  }\href {https://doi.org/10.1038/nature21037} {\bibfield  {journal} {\bibinfo
  {journal} {Nature}\ }\textbf {\bibinfo {volume} {541}},\ \bibinfo {pages}
  {473} (\bibinfo {year} {2017})},\ \Eprint {https://arxiv.org/abs/1608.00446}
  {arXiv:1608.00446} \BibitemShut {NoStop}%
\bibitem [{\citenamefont {Scheucher}\ \emph {et~al.}(2016)\citenamefont
  {Scheucher}, \citenamefont {Hilico}, \citenamefont {Will}, \citenamefont
  {Volz},\ and\ \citenamefont {Rauschenbeutel}}]{scheucher_quantum_2016}%
  \BibitemOpen
  \bibfield  {author} {\bibinfo {author} {\bibfnamefont {M.}~\bibnamefont
  {Scheucher}}, \bibinfo {author} {\bibfnamefont {A.}~\bibnamefont {Hilico}},
  \bibinfo {author} {\bibfnamefont {E.}~\bibnamefont {Will}}, \bibinfo {author}
  {\bibfnamefont {J.}~\bibnamefont {Volz}},\ and\ \bibinfo {author}
  {\bibfnamefont {A.}~\bibnamefont {Rauschenbeutel}},\ }\bibfield  {title}
  {\bibinfo {title} {Quantum optical circulator controlled by a single chirally
  coupled atom},\ }\href {https://doi.org/10.1126/science.aaj2118} {\bibfield
  {journal} {\bibinfo  {journal} {Science}\ }\textbf {\bibinfo {volume}
  {354}},\ \bibinfo {pages} {1577} (\bibinfo {year} {2016})}\BibitemShut
  {NoStop}%
\bibitem [{\citenamefont {Tang}\ \emph {et~al.}(2019)\citenamefont {Tang},
  \citenamefont {Tang}, \citenamefont {Zhang}, \citenamefont {Lu},
  \citenamefont {Zhang}, \citenamefont {Zhang}, \citenamefont {Xia},\ and\
  \citenamefont {Xiao}}]{tang_2019}%
  \BibitemOpen
  \bibfield  {author} {\bibinfo {author} {\bibfnamefont {L.}~\bibnamefont
  {Tang}}, \bibinfo {author} {\bibfnamefont {J.}~\bibnamefont {Tang}}, \bibinfo
  {author} {\bibfnamefont {W.}~\bibnamefont {Zhang}}, \bibinfo {author}
  {\bibfnamefont {G.}~\bibnamefont {Lu}}, \bibinfo {author} {\bibfnamefont
  {H.}~\bibnamefont {Zhang}}, \bibinfo {author} {\bibfnamefont
  {Y.}~\bibnamefont {Zhang}}, \bibinfo {author} {\bibfnamefont
  {K.}~\bibnamefont {Xia}},\ and\ \bibinfo {author} {\bibfnamefont
  {M.}~\bibnamefont {Xiao}},\ }\bibfield  {title} {\bibinfo {title} {On-chip
  chiral single-photon interface: Isolation and unidirectional emission},\
  }\href {https://doi.org/10.1103/PhysRevA.99.043833} {\bibfield  {journal}
  {\bibinfo  {journal} {Phys. Rev. A}\ }\textbf {\bibinfo {volume} {99}},\
  \bibinfo {pages} {043833} (\bibinfo {year} {2019})}\BibitemShut {NoStop}%
\end{thebibliography}%

\clearpage
\newpage

\section*{Appendix}
\section{Traditional waveguide and cavity QED}
\subsection{Coupling a two-level emitter to free-space modes}
\label{app:freespacecoupling}
In order to derive an alternate description to the Jaynes-Cummings Hamiltonian, we start with the situation without optical resonator, where a two-level emitter with resonance frequency $\omega_0$ is coupled to a pair of counter-propagating running waves $\hat a_{1,2}$ which later will constitute the cavity mode and a set of free-space modes $\hat b_{k}$, see also Fig.~\ref{fig:scheme}. We note that this nomenclature is only introduced for consistency with the rest of this manuscript, fundamentally there is no difference between the modes $\hat a_i$ and $\hat b_k$. Following the approach outlined in \cite{Shen2009} the Hamiltonian of the coupled system is 
\small{
\begin{eqnarray}
&&\hat{H}/\hbar=\omega_0\hat \sigma^+\hat \sigma^-\nonumber\\
&&-\sum_{i=1}^2\left(ic\int\limits_{}^{}dx \hat a_i^\dagger(x)\partial_x \hat a_i(x) -V_{a_i}\left[\hat a_i^\dagger(0)\hat \sigma^-+\hat a_i(0)\hat \sigma^+\right]\right)\nonumber\\
&&-\sum\limits_k\left( ic\int\limits_{}^{}dy \hat b_k^\dagger(y)\partial_y \hat b_k(y)-V_{b_k}\left[\hat b_k^\dagger(0)\hat \sigma^-+\hat b_k(0)\hat \sigma^+\right]\right).\label{eq:H_freespace}\nonumber\\
\end{eqnarray}
}
Here, $\partial x=(i\omega_\text{lin}/c+\partial/\partial x)$, where $\omega_\text{lin}$ is the frequency around which the dispersion relation of the light modes can be approximated to be linear and $c$ is the corresponding group velocity which, without loss of generality, we here assume to be the same for all modes. The operators $\hat a_i(x)$ and $\hat b_k(y)$ ($\hat a_i^\dagger(x)$ and $\hat b_k^\dagger(y)$) are the annihilation (creation) operators for photons at position $x$ and $y$ in mode $a_i$ and $b_k$, respectively. The operators $\sigma^+$ and $\sigma^-$ describe the excitation and de-excitation of the atom, respectively. Note that for a compact notation, we assign each mode its own coordinate system which is chosen such that the atom is always located at the position $x=0$ and $y=0$ in the respective mode and that the light always propagates in positive $x$- and $y$-direction. All integrals until specified otherwise extend from $-\infty$ to $\infty$. 
The constants $V_{a_i}$ and $V_{b_k}$ describe the interaction between the atom and the respective modes. Here and in the following, we assume, without loss of generality, these coupling constants to be real. In the basis used for our Hamiltonian, the general quantum state for the case of a single excitation is given by
\begin{eqnarray}
|\psi\rangle&=&\Big[\phi_0 \hat \sigma^+ +\sum_{i=1}^2\int dx \phi_{a_i}(x)\hat a_i^\dagger(x)\nonumber\\
&&+\sum_k\int dy \phi_{b_k}(y)\hat b_k^\dagger(y)\Big]|0\rangle,
\end{eqnarray}
with the complex amplitudes $\phi_0$, $\phi_{a_i}(x)$ and $\phi_{b_{k}}(y)$. The state $|0\rangle=|g\rangle\otimes|0\rangle$ corresponds to the ground state of the system, with the atomic ground state $|g\rangle$ and the field vacuum state $|0\rangle$. In the following we are interested in steady state solutions for which all propagating fields must be of the form 
\begin{eqnarray}
\phi_j(z)=e^{ikz}\left\{
\begin{array}{ll}
\phi_j^\text{in} & \text{for } z=(x,y)<0\\
\phi_j^\text{out} & \text{for } z=(x,y)>0
\end{array} \right.,
\end{eqnarray}
where $k$ is the wavenumber of the solution in the respective mode. Using this ansatz for the time-independent Schr\"odinger equation $\hat H|\psi\rangle=\varepsilon |\psi\rangle$ and comparing the coefficients for the different parts of the wavefunction, we obtain a set of linear equations for the steady state solutions
\begin{eqnarray}
0&=&\int dy \left[(\omega_p-\frac{\varepsilon}{\hbar})\phi_{b_k}(y)-ic\frac{\partial}{\partial y}\phi_{b_k}(y)+\delta(y)V_{b_k}\phi_0\right]\nonumber\\
\\
0&=&\int dx \left[ (\omega_p-\frac{\varepsilon}{\hbar})\phi_{a_i}(x)-ic\frac{\partial}{\partial x}\phi_{a_i}(x)+\delta(x)V_{a_i}\phi_0\right]\nonumber\\
\\
0&=&(\omega_0-\frac{\varepsilon}{\hbar})\phi_0+\sum_{i=1}^2 V_{a_i}\phi_{a_i}(0)+\sum_k V_{b_k}\phi_{b_k}(0),
\end{eqnarray}
where $\omega_p=ck-\omega_\text{lin}$ is the eigenfrequency of the solution.
In the case where we probe the system via mode $\hat a_1$ and all other incoming modes are unpopulated, i.e., $\phi^\text{in}_{a_2}=\phi^\text{in}_{b_k}=0$, we obtain 
\begin{eqnarray}
\phi_{a_1}^\text{out}/\phi^\text{in}_{a_1}&=& 1-\frac{2V_{a_1}^2}{n} \equiv t_\textrm{at}\label{eq:t}\\
\phi_{a_2}^\text{out}/\phi^\text{in}_{a_1}&=&- \frac{2 V_{a_1}V_{a_2}}{n} \equiv r_\textrm{at}\label{eq:r}\\
\phi_{b_k}^\text{out}/\phi^\text{in}_{a_1}&=& -\frac{2 V_{a_1}V_{b_k}}{n} \\
\phi_0/\phi^\text{in}_{a_1}&=& - \frac{2ic V_{a_1}}{n} \label{eq:atomic_exictation}
\end{eqnarray}
with
\begin{equation} 
n=V_{a_1}^2+V_{a_2}^2+\sum_k V_{b_k}^2+2ic(\omega_0-\omega_p).
\end{equation} 
Here, $r_\textrm{at}$ and $t_\textrm{at}$ are the reflection and transmission coefficient of the atom for light propagating along the mode $\hat a_1$. Note that Eq.~(\ref{eq:atomic_exictation}) is a Lorentzian that describes the atomic excitation as a function of atom-light detuning $(\omega_0-\omega_p)$. If the modes $\hat a_i$ and $\hat b_k$ span the whole range of optical modes, the width of the Lorentzian must be equal to the total atomic field decay rate $\gamma$ and we get
\begin{equation}
V_{a_1}^2+V_{a_2}^2+\sum_k V_{b_k}^2=2c\gamma.
\end{equation}
If we define $\beta_{b_k}$ ($\beta_{a_i}$) as the fraction of photons that is scattered into a given mode $\hat b_k$ ($\hat a_i$), it follows from Eqs.~(\ref{eq:t})-(\ref{eq:atomic_exictation}) that it is proportional to $V_{b_k}^2$ ($V_{a_i}^2$) and we obtain 
\begin{eqnarray}
V_{a_i}^2=2c\beta_{a_i} \gamma\\
V_{b_k}^2=2c\beta_{b_k} \gamma
\end{eqnarray}	
with $\sum_k\beta_{b_k}+\sum_i\beta_{a_i}=1$. Typically one is not interested in the time evolution of the free-space modes $\hat b_{k}$ and in the case where no light is incident from these modes, their effect on the time evolution of the atom and the modes $\hat a_i$ in Eqs.~(\ref{eq:t})-(\ref{eq:atomic_exictation}) only enters in the nominator as
\begin{eqnarray}
n&=&2c\Big[\gamma(\beta_{a_1}+\beta_{a_2})+\overbrace{\gamma\sum\limits_k \beta_{k}}^{\gamma_l}+i(\omega_0-\omega_p)\Big]\nonumber\\
&=&2c\Big[\gamma(\beta_{a_1}+\beta_{a_2})+i(\omega_0-i\gamma_l-\omega_p)\Big].\nonumber\\
\end{eqnarray}
Thus, the solutions for the modes $\hat a_i$ are identical to the solutions of a simplified Hamiltonian, which one obtains from  Eq.~(\ref{eq:H_freespace}) when dropping all terms containing $\hat b$ and $\hat b^\dagger$ and adding the term $-i\gamma_l \hat\sigma^+\hat\sigma^-$. We apply this replacement throughout the manuscript whenever the evolution of the free-space modes are not of interest for the problem.
Using the above definitions, we can simplify Eqs.~(\ref{eq:t})-(\ref{eq:atomic_exictation}) and obtain 
\begin{eqnarray}
\phi_{a_1}^\text{out}/\phi^\text{in}_{a_1}&=&t= 1-2\frac{\beta_{a_1}\gamma}{\gamma+i\Delta} \\
\phi_{a_2}^\text{out}/\phi^\text{in}_{a_1}&=&r=-2\frac{\sqrt{\beta_{a_1}\beta_{a_2}}\gamma}{\gamma+i\Delta} \\
\phi_{b_k}^\text{out}/\phi^\text{in}_{a_1}&=& -2\frac{\sqrt{\beta_{a_1}\beta_{b_k}}\gamma}{\gamma+i\Delta} \\
\phi_0/\phi^\text{in}_{a_1}&=& - i\frac{ \sqrt{2c\beta_{a_1}\gamma}}{\gamma+i\Delta}.
\end{eqnarray}

\subsection{Solutions of the Jaynes-Cummings model}\label{app:JC}
Now, we consider the case of a single atom coupled to a single mode of a Fabry-P\'erot resonator with coupling strength $g$ in the Jaynes-Cummings description. The cavity consists of two mirrors via which the cavity mode $\hat a$ couples to the propagating outside field modes $\hat m_1(y)$ and $\hat m_2(y)$, respectively. In addition, the atom also couples to a set of free-space light field modes $\hat b_k(y)$, see Fig.~\ref{fig:scheme}. Including these external probe fields, the generalized Jaynes-Cummings Hamiltonian describing the probed atom-cavity system can be written as
\begin{eqnarray}
&&\hat{H}/\hbar=\omega_0 \hat{\sigma}^+ \hat{\sigma}^- + \omega_a \hat{a}^{\dagger} \hat{a} + g( \hat{a}^{\dagger} \hat{\sigma}^- + \hat{a} \hat{\sigma}^+)+\hat U_\text{cpl}\nonumber
\end{eqnarray}
with the coupling Hamiltonian
\begin{eqnarray}
&&\hat U_\text{cpl}=-i\kappa_0 \hat{a}^{\dagger} \hat{a} \nonumber\\
&&-\sum_k \left(ic\int dy\;\hat b_k^\dagger(y)\partial_y \hat b_k(y)-\Big[ V_{b_k}(\hat b_k(0)\hat \sigma^++\hat b_k^\dagger(0)\hat \sigma^-)\Big]\right)\nonumber\\
&&-\sum_{k=1}^{2}\left( ic\int dy\; \hat m_k^\dagger(y)\partial_y\hat m_{k}(y)-V_{m_k}\left[\hat m_k(0)\hat a^\dagger+\hat m_{k}^\dagger(0) \hat a\right]\right).\nonumber\\
\end{eqnarray}
The operators $\hat m_1(y)$, $\hat m_2(y)$, $\hat b_k(y)$ ($\hat m_1^\dagger(y)$, $\hat m_2^\dagger(y)$, $\hat b_k^\dagger(y)$) are the annihilation (creation) operators for photons incident onto the cavity mirrors and the atoms, respectively. The free-space--cavity (free-space--atom) coupling is given by $V_{m_k}=\sqrt{2c\kappa_k}$ ($V_{b_k}=\sqrt{2c\beta_{b_k}\gamma}$), see previous section, where $\kappa_k$ ($\beta_{b_k}\gamma$) is the decay rate of the resonator (atom) through mirror $k$ (into the free-space mode $k$). In addition to this, we also included the loss rate $\kappa_0$ to describe additional losses of the resonator, e.g., due to absorption or scattering losses in the mirrors. As we are not interested in the time evolution of the free-space modes $\hat b_k$ except for one mode $\hat b^\dagger$ that is used to excite the atom, we can describe their effect on the atom-resonator system by using the simplified coupling Hamiltonian (see previous section):
\begin{eqnarray}
\hat U_\text{cpl}= -i\gamma_l \hat{\sigma}^+ \hat{\sigma}^- -i\kappa_0\hat{a}^{\dagger} \hat{a} +\hat U_\text{probe}
\end{eqnarray}
with 
\begin{eqnarray}
&&\hat U_\text{probe}=\nonumber\\
&&-\left(ic\int dy\;\hat b^\dagger(y)\partial_y \hat b(y)-\Big[ V_{b}(\hat b(0)\hat \sigma^++\hat b^\dagger(0)\hat \sigma^-)\Big]\right)\nonumber\\
&&-\sum_{k=1}^{2}\left( ic\int dy\; \hat m_k^\dagger(y)\partial_y\hat m_{k}(y)-V_{m_k}\left[\hat m_k(0)\hat a^\dagger+\hat m_{k}^\dagger(0) \hat a\right]\right),\nonumber\\ \label{eq:Uprobe}
\end{eqnarray}
where $\gamma_l=(1-\beta_{a_1}-\beta_{a_2}-\beta_b)\gamma$. Note that in the following, we either neglect coupling of the emitter to external modes or consider the case, where the external mode $\hat b$ is only weakly coupled to the emitter, i.e., $\beta_b\ll1$. Thus, we  get $\gamma_l=(1-\beta_{a_1}-\beta_{a_2})\gamma$. The operator $\hat U_\text{cpl}$ describes situations in CQED where either the atom is probed via an external light field or the resonator is probed by sending light onto the incoupling mirrors. Similar to the free-space situation, the general quantum state of the system for a single excitation is then given by
\begin{eqnarray}
|\psi\rangle&=&\Big[\phi_0 \hat\sigma^+ +\phi_a \hat a^\dagger +\int dy \phi_{b}(y) \hat b^\dagger(y)\nonumber\\
&&+\sum_{k=1}^{2}\int dy \phi_{m_k}(y)\hat m_k^\dagger(y)\Big]|0\rangle,
\end{eqnarray}
where the coefficients $\phi_0$, $\phi_{a}$, $\phi_{m_{1,2}}(y)$ and $\phi_{b}(y)$ are complex amplitudes. 
Using the same method as in the previous section, we obtain for the steady state a set of linear equations for the amplitudes which can be analytically solved. 

A standard situation of CQED is the case where the cavity is probed through mirror $m_1$ with amplitude $\phi_{m_1}^\text{in}=\phi_{in}$ and all other incident field modes are empty, i.e., $\phi^\text{in}_{m_2}=\phi^\text{in}_{b}=0$ . Solving this set of equations for this case we obtain the steady state amplitudes 
\begin{eqnarray}
\frac{\phi_a}{\phi_{in}}&=&-i\sqrt{2c\kappa_1}\frac{(\gamma_{l}+i\Delta_0)}{g^2+(\gamma_l+i\Delta_0)(\kappa_l+i\Delta_a)}\label{eq:JC1}\\
\frac{\phi_0}{\phi_{in}}&=&-\sqrt{2c\kappa_1}\frac{g}{g^2+(\gamma_l+i\Delta_0)(\kappa_l+i\Delta_a)}\label{eq:JC2}\\
\frac{\phi_{ref}}{\phi_{in}}&=&1-i\sqrt{\frac{\kappa_1}{c}}\frac{\phi_a}{\phi_{in}}\\
\frac{\phi_{trans}}{\phi_{in}}&=&-i\sqrt{\frac{\kappa_2}{c}}\frac{\phi_a}{\phi_{in}},
\end{eqnarray}
where we introduced the atom-probe (resonator-probe) detuning $\Delta_0=\omega_0-\omega_p$ ($\Delta_a=\omega_a-\omega_p$), the total resonator loss rate $\kappa_l\equiv\kappa_0+\kappa_1+\kappa_2$ and $\phi_{ref}$ and $\phi_{trans}$ are the fields reflected from and transmitted through the cavity, see Fig.~\ref{fig:scheme}. 
From the first two expressions one obtains the excited state amplitude of the atom as a function of the cavity field as
\begin{equation}
\phi_0/\phi_a=-i\frac{g}{\gamma_l+i\Delta_0}. \label{eq:Gamma}
\end{equation}

In the same way, we can calculate the steady state solutions when probing the system via the external mode $\hat b$ with the amplitude $\phi_{in}$ while all other incoming fields are zero. We obtain
\begin{eqnarray}
\frac{\phi_a}{\phi_{in}}&=&-\sqrt{2c\beta_b\gamma}\frac{g}{g^2+(\gamma_l+i\Delta_0)(\kappa_l+i\Delta_a)}\\
\frac{\phi_0}{\phi_{in}}&=&-i\sqrt{2c\beta_b\gamma}\frac{\kappa_l+i\Delta_a}{g^2+(\gamma_l+i\Delta_0)(\kappa_l+i\Delta_a)}\\
\frac{\phi_{ref}}{\phi_{in}}&=&-i\sqrt{\frac{\kappa_1}{c}}\frac{\phi_a}{\phi_{in}}\\
\frac{\phi_{trans}}{\phi_{in}}&=&-i\sqrt{\frac{\kappa_2}{c}}\frac{\phi_a}{\phi_{in}},
\end{eqnarray}
where the fields $\phi_{trans}$ and $\phi_{ref}$ correspond to the same two output fields of the cavity as before, see Fig.~\ref{fig:scheme}. Similar as in the case where we probe the cavity via the mirror we obtain for the cavity field excitation a function of the atom excitation
\begin{equation}
\phi_a/\phi_0=-i\frac{g}{\kappa_l+i\Delta_a}. \label{eq:Kappa}
\end{equation}

\subsection{Effective loss rates in the JC model}
\label{app:lossrate}
From the steady state population of the atomic excited state and the cavity field in Eqs.~(\ref{eq:Gamma}) and (\ref{eq:Kappa}) we can calculate the corresponding effective loss rates $\Gamma$ and $K$ that on resonance are given by
\begin{eqnarray}
\Gamma&=&\frac{|\phi_0|^2}{|\phi_{a}|^2}\gamma_l=\frac{\rho_e}{\rho_a}\gamma_l=g^2/\gamma_l \\
K&=&\frac{|\phi_a|^2}{|\phi_{0}|^2}\kappa_l=\frac{\rho_a}{\rho_e}\kappa_l=g^2/\kappa_l.
\end{eqnarray}
Here, $\Gamma$ describes the emitter-induced amplitude decay rate of the cavity mode and $K$ describes the cavity-induced dipole decay rate of the emitter. 
Surprisingly, these effective decay rates diverge once the bare coupling rates to the environment $\gamma_l$ and $\kappa_l$ approach zero.
Consequently, these effective loss rates will be the fastest rates in the system which contradicts the assumption of a global cavity field in the Jaynes-Cummings Hamiltonian, as the rate of change is now no longer small with respect to the inverse of the characteristic time scale of the resonator field $\tau_c^{-1}=c/L=\nu_{\text{fsr}}$ or the emitter $\tau_0^{-1}\propto c/d$. Here, $L$ and and $d$ are the effective roundtrip lengths of the resonator and emitter, respectively. Per definition, in CQED, the spatial extension of the resonator, $L$, is larger - typically by many orders of magnitude - than that of the emitter, $d$. Consequently, the spatial extension of the emitter can in most cases be neglected. Thus in this manuscript we limit our discussion to the behavior of the coupled system with respect to $ \gamma_{l} $.
We note however that, in particular in circuit CQED system, the spatial extension of the emitter can also be significant. If this is the case, the emitter itself cannot be treated any more as point-like and one has to describe the excitation transport inside the emitter via position dependent operators, similar to our approach for the cavity field.

\section{Cascaded model for CQED}
\subsection{Emitter-cavity Hamiltonian}
\label{app:cascHam}
The problems in the JC Hamiltonian arise from the assumption of a global cavity field that can be described by a single set of operators $\hat a$ and $\hat a^\dagger$ which is only valid for negligible change of the cavity field in a single round trip. In order to derive an alternative description of light-matter interaction that is not subject to these limits, we explicitly take into account the spatial evolution of the circulating light field in the cavity by replacing the global field operators $\hat a$ and $\hat a^\dagger$ by their position dependent counterparts $\hat a(x)$ and $\hat a^\dagger(x)$, respectively, where $ x $ is the axial position in the resonator. In analogy to the JC description, we derive a Hamiltonian for the coupled system which is given by
\begin{eqnarray}
&&\hat{H}/\hbar=\omega_0\hat \sigma^+\hat \sigma^-\nonumber\\
&&-ic\int\limits_{0}^{L/2}dx \hat a^\dagger(x)\partial_x \hat a(x)-ic\int\limits_{L/2}^{L}dx' \hat a^\dagger(x')\partial_{x'} \hat a(x')\nonumber\\ 
&&+V_{a_1}(\hat a^\dagger(x_a)\hat \sigma^-+\hat a(x_a)\hat \sigma^+)\nonumber\\
&&+V_{a_2}(\hat a^\dagger(L-x_a)\hat \sigma^-+\hat a(L-x_a)\hat \sigma^+)\nonumber\\
&&+c\left[ \hat U_{m_1}(L,0)+\hat U_{m_2}(L/2,L/2)\right]+\hat{U}_\text{cpl}.
\label{eq:ourJCfull}
\end{eqnarray}
Here, 
\begin{eqnarray}
\hat{U}_{m_k}(x,x')&=& -ir_k\hat a(x)\hat a^\dagger(x')-t_k\hat a(x)\hat m_k^\dagger(0)\nonumber\\
&&-ir_k \hat m_k'(0)\hat m_k^\dagger(0)-t_k \hat m_k'(0)\hat a^\dagger(x')
\end{eqnarray}
is the beamsplitter operator with the field transmission (reflection) coefficient $t_k$ ($r_k$) that describes the cavity-mode-field coupling at mirror $k$. This description has to be introduced since, in contrast to the JC approximation in the previous chapter, the cavity is no longer considered to be a point-like system. The fields $\hat m_k$ separate into an incoming field ($\hat m_k'(x)$ for $x\in[-\infty,0[$) and an outgoing field ($\hat m_k(x)$ for $x\in[0,\infty]$) that are connected via the coupling mirror. The coupling Hamiltonian is then given by 
\begin{eqnarray}
&&\hat{U}_\text{cpl}/\hbar=\nonumber\\
&&-ic\sum\limits_{k=1}^2\Big[ 
\int\limits_{-\infty}^0dy\;\hat m_k'^\dagger(y)\partial_y \hat m'_k(y)+\int\limits_0^{\infty}dy\;\hat m_k^\dagger(y)\partial_y \hat m_k(y)\Big]\nonumber\\
&&-\sum\limits_k \Big( ic\int\limits_{}^{}dy \hat b_k^\dagger(y)\partial_y \hat b_k(y)-V_{b_k}\left[\hat b_k^\dagger(0)\sigma^-+b_k(0)\hat \sigma^+\right]\Big)\nonumber\\
\end{eqnarray}
and analogous to the previous section by explicitly considering one external excitation mode of the emitter $\hat b$ we can simplify the coupling Hamiltonian to
\begin{eqnarray}
&&\hat{U}_\text{cpl}/\hbar=-i\gamma_l \hat{\sigma}^+ \hat{\sigma}^-\nonumber\\
&&-ic\sum\limits_{k=1}^2\Big[ 
\int\limits_{-\infty}^0dy\;\hat m_k'^\dagger(y)\partial_y \hat m'_k(y)+\int\limits_0^{\infty}dy\;\hat m_k^\dagger(y)\partial_y \hat m_k(y)\Big]\nonumber\\
&&-\left(ic\int dy\;\hat b^\dagger(y)\partial_y \hat b(y)-\Big[ V_{b}(\hat b(0)\hat \sigma^++\hat b^\dagger(0)\hat \sigma^-)\Big]\right).\nonumber\\\label{eq:Ucpl}
\end{eqnarray}

In this picture, we describe the resonator mode as a running wave circulating in the cavity. For a Fabry-P\'erot resonator, the light will thus interact twice with the atom at the position $x=x_{a}$ and $x=L-x_{a}$ with the coupling strength $V_{a_1}=\sqrt{2c\beta_{a_1}\gamma}$ and $V_{a_2}=\sqrt{2c\beta_{a_2}\gamma}$, respectively. In Eq.~(\ref{eq:Ucpl}) we trace over the freely-propagating fields and the emitter's scattering rate into the environment, $\gamma_l$, is given by  $\gamma_l=\gamma(1-\beta_{a_1}-\beta_{a_2})$.
We note that Eq.~(\ref{eq:ourJCfull}) has no explicit term describing cavity losses such as absorption in the mirrors. However, these can be straightforwardly included by, e.g., introducing mirror reflection and transmission coefficients $r_i$ and $t_i$ for which $|r_i|^2+|t_i|^2<1$.

\section{Steady state solutions}
\label{app:steadystate}
In order to solve the emitter-cavity Hamiltonian in Eq.~(\ref{eq:ourJCfull}), we follow the same approach as before and calculate the steady state solutions for the case where we continuously probe the system via the first mirror with an input field with amplitude $\phi_{in}$. In the following we consider two cases: the case where the atom couples equally to both propagation directions of the cavity field, i.e., $\beta_{a_1}=\beta_{a_2}$ which corresponds to the situation in a typical Fabry-P\'erot resonator. In addition we consider the case where the atom-resonator interaction is fully chiral \cite{lodahl_chiral_2017}, i.e., $\beta_{a_2}=0$ and $\beta_{a_1}>0$, a situation that can occur when an emitter couples to a resonator field with very strong field gradients that exits, e.g., in whispering-gallery-mode resonators \cite{junge2013} or ring resonators with nanoscale waveguides \cite{johnson_observation_2019}.

\subsection{Fabry-P\'erot resonator}
\label{app:FP}
Here, we consider the case where an emitter couples equally to both propagation directions of the cavity field, i.e., $\beta_{a_1}=\beta_{a_2}\equiv\beta/2$. Solving the time-independent Schr\"odinger equation, we then derive the amplitudes of the 4 different cavity fields and the atomic excitation which are given by
\begin{eqnarray}
\phi_1/\phi_{in}&=&-it_1\frac{e^{i\alpha}r_2\tilde\beta-1}{N}\label{eq:our_solution1}\\
\phi_2/\phi_{in}&=&-it_1\frac{ \tilde\beta -1}{N}\\
\phi_3/\phi_{in}&=& it_1r_2\frac{\tilde\beta -1}{N}\\
\phi_4/\phi_{in}&=&it_1\frac{r_2(2\tilde\beta -1)-e^{-i\alpha}\tilde\beta}{N}\label{eq:our_solution4}\\
\phi_0/\phi_{in}&=&-t_1e^{-i\frac{ \Delta_a }{2 \nu_{\text{fsr}}}}\sqrt{\frac{\tilde\beta c}{\gamma +i \Delta_0 }}\frac{ (e^{i\alpha/2}r_2-e^{-i\alpha/2})} {N}
\label{eqn:our_solutions}
\end{eqnarray}
with the common denominator
\begin{equation}
N=1-e^{-i\frac{ \Delta_a }{\nu_{\text{fsr} }}}e^{-i\alpha} r_1 \tilde\beta -e^{i\alpha}r_2\tilde\beta + e^{-i\frac{ \Delta_a }{\nu_{\text{fsr} }}}r_1r_2(2\tilde\beta-1).
\end{equation}
The amplitude reflection and transmission of the whole resonator-atom system are given by
\begin{eqnarray}
\phi_{ref}/\phi_{in}&=&i r_1+t_1\phi_4/\phi_{in}\\
\phi_{trans}/\phi_{in}&=&t_2\phi_2/\phi_{in}.
\end{eqnarray}
Here, we used the identity
\begin{equation}
\exp(ikL)=\exp(-i\Delta_a/\nu_{\text{fsr} })
\end{equation}
with the wavenumber $k$ and introduced the detuning-dependent $\beta$ factor
\begin{equation} 
\tilde\beta=\frac{\beta}{1+i\Delta_0/\gamma},
\end{equation}
as well as the phase difference $\alpha$ of the two running wave modes at the atom's position via
\begin{equation}
\alpha=k(L-2x_a)=\alpha_0-\Delta_a\nu_\text{fsr}(1-2x_a/L) \label{eq:alpha}.
\end{equation}
The phase $\alpha_0$ describes the position of the atom in the standing wave mode of the cavity, where maximum (zero) coupling is reached when $\alpha_0=(2n+1)\pi$ ($\alpha_0=2\pi n$).

\subsection{Ring resonator with chiral coupling} \label{app:chiral}
\begin{figure}[tb]
	\includegraphics[width=0.9\columnwidth]{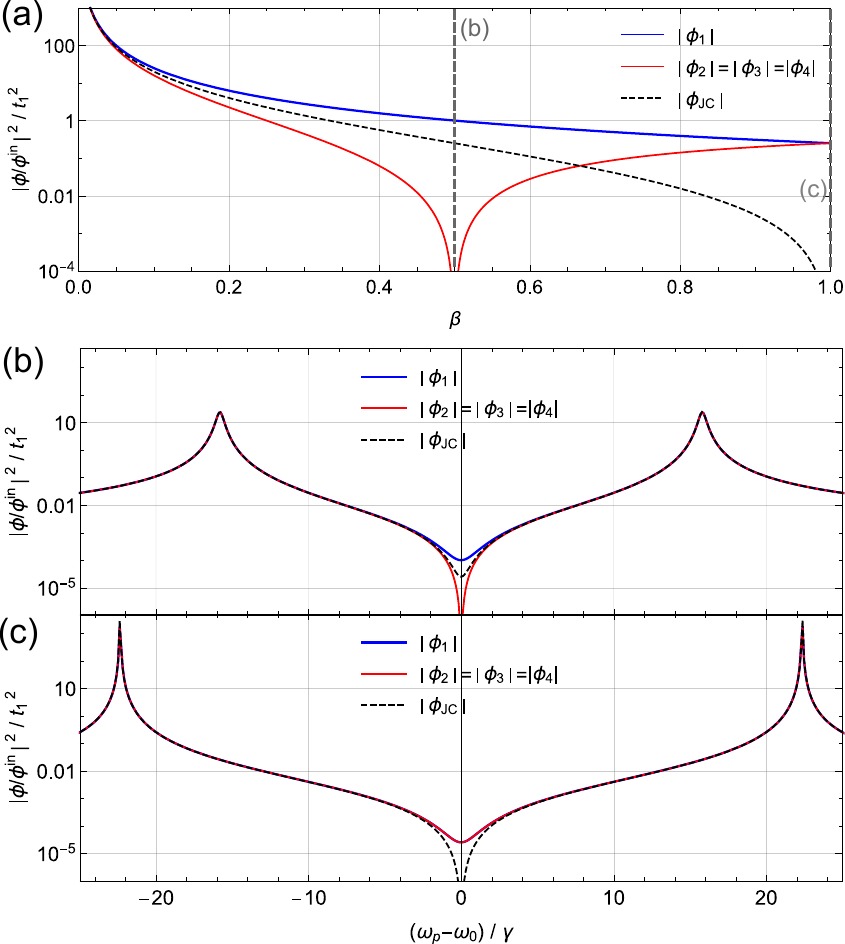}	
	\caption{(a) On-resonance field strength in the ring resonator as a function of atom channeling efficiency $\beta$ for a high-finesse resonator for an atom chirally coupled to one of the two ring-resonator modes. The solid lines indicate the prediction of our cascaded model and the dashed line that of the JC model. (b) and (c) Excitation of the different cavity fields as a function of light-resonator detuning for $\beta=1/2$ and $\beta=1$, respectively, assuming $t_1^2=t_2^2=1-r_1^2=1-r_2^2=10^{-4}$ and $\nu_{\text{fsr} }=250\gamma$. As the cavity field for a ring resonator is given by a running wave, the atom-resonator coupling strength is independent of the longitudinal position of the atom. Consequently, all predictions of the absolute field values are independent of $\alpha$.}
	\label{fig:betaC}
\end{figure}

Another physical situation that for small $\beta$ can be described by the JC model is a single atom that is chirally coupled to an optical ring resonator \cite{junge2013,lodahl_chiral_2017}. In this situation, the atom only couples to one of the two counter-propagating resonator modes and one again obtains the situation of a two-level emitter coupled to a single mode of the electromagnetic field. 
From the Hamiltonian in Eq.~(\ref{eq:ourJCfull}), we can straightforwardly derive the predictions for this case, by assuming full chiral coupling, i.e., $V_{a_2}=0$ and $V_{a_1}\equiv V_a>0$. Solving the Hamiltonian in the same way as in Appendix~\ref{app:JC} yields for the steady amplitudes
\begin{eqnarray}
\phi_1/\phi_{in}&=&-it_1\frac{1}{N}\\
\phi_2/\phi_{in}&=&-it_1\frac{1-2\tilde\beta}{N}\\
\phi_3/\phi_{in}=\phi_4/\phi_{in}&=&-it_1r_2\frac{1-2\tilde\beta}{N}\\
\phi_0/\phi_{in}&=&-t_1e^{-i\frac{\Delta_a}{2\nu_{\text{fsr}}}}\sqrt{\frac{2c\tilde\beta}{\gamma+i\Delta_0}}\frac{e^{-i\frac{\alpha}{2}}}{N}
\end{eqnarray}
with the common denominator
\begin{equation}
N=1+e^{-i\frac{\Delta_0}{\nu_{\text{fsr}}}}r_1r_2(2\tilde\beta-1).
\end{equation}
Figure~\ref{fig:betaC} shows the dependence of the field amplitudes $|\phi_1|...|\phi_4|$ on $\beta$ and compares it to the predictions of the Jaynes-Cummings model. The curves show a similar behavior as the Fabry-P\'erot case discussed in the main part of the manuscript but here the situation for which the atom acts as a perfect absorber is reached for $\beta=1/2$.
Figure~\ref{fig:betaC} (b) and (c) shows the corresponding Rabi-spectra for the channeling efficiencies $\beta=1/2$ and $\beta=1$, respectively. 

\section{Relation of the parameters of the different models}\label{app:defg}
In the different models presented above, we use a variety of parameters to define the interaction between the atom and the optical modes. In order to quantitatively compare the predictions of the JC model with our cascaded model, we have to relate these quantities. For the case of a point-like interaction between an atom (cavity mode) with an external optical mode, the coupling is given by $V_{b_k}$ ($V_{m_k}$) which relate to the decay rate into the respective mode via
\begin{eqnarray}
	V_{m_{k}}&=&\sqrt{2c\kappa_{k}}\\
	V_{b_k}&=&\sqrt{2c\gamma_{b_k}}=\sqrt{2c\gamma\beta_{b_k}}.
\end{eqnarray}
The intrinsic resonator decay rate $\kappa_0$ and the decay rate through the mirrors $\kappa_{1,2}$ are related to the intrinsic roundtrip loss $l_{0}$ and mirror transmissions $t_{1,2}$ via 
\begin{eqnarray}
	l_0=\sqrt{1-t_{0}^2}&=&\sqrt{\frac{2\kappa_0}{\nu_{\text{fsr}}}}\label{eq.t0}\\
	t_k=\sqrt{1-r_k^2}&=&\sqrt{\frac{2\kappa_k}{\nu_{\text{fsr}}}},\label{eq.t1}
\end{eqnarray}
where $ t_0 $ is the roundtrip transmission (excluding loss due to mirror transmission).

In the JC model, the atom-resonator coupling strength is given by $g$ whereas in our cascaded model it is given by $V_{a_1}$ and $V_{a_2}$ and the phase difference $\alpha$ of the propagating resonator mode arriving from the left and the right at the atom, see Eq.~(\ref{eq:alpha}). For symmetric coupling ($V_{a_1}=V_{a_2}=V_a$) which corresponds to the typical situation in a Fabry-P\'erot resonator, one can define an effective coupling strength to the standing wave cavity mode which is given by 
\begin{equation}
	V_\text{eff}=V_a(1-e^{i\alpha})=-2ie^{i\alpha/2}\sin(\alpha/2)\sqrt{c\beta\gamma},
	\label{eq:definition_g1}
\end{equation}
where $\beta=2\beta_{a_1}=2\beta_{a_2}$ defines the decay of the atom into the standing wave resonator mode.
In the JC model, the coupling strength $g$ describes the interaction between the atom and the photon state in the whole cavity. Thus, in order to relate $V_\text{eff}$ to $g$, we have to compare situations with the same intracavity photon number in both models. We thus require
\begin{equation}
	\int_0^L dx\langle \hat a^\dagger(x)\hat a(x)\rangle=n_\text{cav}=\langle \hat a^\dagger \hat a\rangle.
\end{equation}
For small $\beta$, $\langle \hat a^\dagger(x)\hat a(x)\rangle$ is independent of $x$ and, thus, we get the relation $g=|V_\text{eff}|/\sqrt{L}$ and with $\nu_\text{fsr}=c/L$ this simplifies on resonance to
\begin{equation}
	g^{\text{FP}}=2|\sin(\alpha_0/2)|\sqrt{\beta\gamma\nu_{\text{fsr}}}.
	\label{eq:definition_g2}
\end{equation}
This relates the coupling constant $ g^{\text{FP}} $ in the Jaynes-Cummings model to the $ \beta $ factor in our cascaded model.

For the situation where the emitter is located at an anti-node of the mode and thus maximally coupled to the resonator, $\alpha_0=(2n+1)\pi$ ($n\in\mathcal R$), we obtain
\begin{equation} 
	g^{\text{FP}}_{\text{max}}=2\sqrt{\beta\gamma\nu_{\text{fsr}}}.
\end{equation}
Similarly, for the case of an atom that is chirally coupled to the mode of a ring resonator, we obtain independent of position
\begin{equation}
	g^{\text{ring}}_{\text{max}}=\sqrt{2\beta\gamma\nu_\text{fsr}}.\label{eq:gring}
\end{equation}

\subsection{Solutions of the Jaynes-Cummings model}\label{app:JC_solution}
Using the relations in section \ref{app:defg}, we can now express the steady state solutions of the JC model in the parameters of our cascaded approach. This allows us to directly compare the predictions of the two models. In the following, using Eqs.~(\ref{eq.t0}), (\ref{eq.t1}) and (\ref{eq:definition_g2}) we 
express the solutions of the JC Hamiltonian in Eqs.~(\ref{eq:JC1}) and (\ref{eq:JC2}) in terms of $\beta$ and transmission factors $t_{1,2}$ and obtain for a Fabry-P\'erot cavity
\begin{eqnarray}
\phi_a/\phi_{in}&=&-i t_1\frac{1-\tilde\beta}{N_{\textrm{JC}}^\textrm{FP}}\label{eq:compareJC1}\\
\phi_0/\phi_{in}&=&-2t_1\frac{|\sin(\alpha_0/2)|}{N_{\textrm{JC}}^\textrm{FP}}\cdot\sqrt{\frac{c\tilde\beta}{(\gamma+i\Delta_0)}}\label{eq:compareJC2}
\end{eqnarray}
with the common denominator
\begin{equation}
N_{\textrm{JC}}^\textrm{FP}=(1-\tilde\beta)\left(\frac{l_{tot}^2}{2}+\frac{i\Delta_a}{\nu_\text{fsr}}\right)+4\tilde\beta\sin^2(\alpha_0/2)
\end{equation}
where $l_{tot}=l_0^2+t_1^2+t_2^2$ is the total roundtrip loss of the resonator. 

Similarly, using Eqs.~(\ref{eq.t0}), (\ref{eq.t1}) and (\ref{eq:gring}) we obtain for the case of a chiral ring resonator 
\begin{eqnarray}
\phi_a/\phi_{in}&=&-i t_1\frac{1-\tilde\beta}{N_{\textrm{JC}}^\textrm{ring}}\label{eq:compareJC1r}\\
\phi_0/\phi_{in}&=&2\sqrt{2}t_1\frac{1}{N_{\textrm{JC}}^\textrm{ring}}\cdot\sqrt{\frac{c\tilde\beta}{(\gamma+i\Delta_0)}}\label{eq:compareJC2r}
\end{eqnarray}
with the common denominator
\begin{equation}
N_{\textrm{JC}}^\textrm{ring}=(1-\tilde\beta)\left(\frac{l_{tot}^2}{2}+\frac{i\Delta_a}{\nu_\text{fsr}}\right)+2\tilde\beta\,.
\end{equation}

In the limit $\tilde\beta\ll1$ and for not too large light-cavity detunings $\Delta_a\ll\nu_{\text{fsr}}$, the solutions for the four cavity fields Eqs.~(\ref{eq:our_solution1})-(\ref{eq:our_solution4}) in our cascaded model are approximately identical and agree with the prediction of the Jaynes-Cummings model in Eq.~(\ref{eq:compareJC1}). The same is true for the prediction of the atomic excitation by the two models in Eq.~(\ref{eq:compareJC2}) and (\ref{eq:compareJC2r}). So for the case of an emitter that is weakly coupled to an optical resonator, $\beta\ll1$, the two models give identical predictions and the Jaynes-Cummings model is a good approximation for this case. 

\section{Position dependence of the atom-resonator interaction}\label{chapter:alphadep}
For the case of an atom coupled to a Fabry-P\'erot resonator, the light-matter interaction depends on the position $x_a$ of the atom inside the resonator, which in our equations is included in the angle $\alpha$ defined in Eq.~(\ref{eq:alpha}). Interestingly, in this equation, in general $\alpha$ is a function of the light-resonator detuning $\Delta_a$. This means, that when measuring the atom-resonator spectrum with external light, the interference condition of the two running wave contributions that arrive at the atom vary with the probe frequency. Thus, we expect a modification of the atom-light coupling. For $\nu_{\text{fsr} }\gg\gamma$, in the strong coupling regime ($g\gg\gamma,\kappa$) and for zero atom-resonator detuning $\omega_a=\omega_0$, we obtain the detuning $\Delta_{a,max}$ for which the resonator fields reach their maximum value
\begin{equation}
\Delta_{a,max}\approx\pm g^{\text{FP}} - \beta\gamma \cdot \sin\alpha_0 \cdot \left(\frac{2x_a}{L}-\frac{1}{2}\right). \label{eq:gshift}
\end{equation}
One sees that in contrast to the JC model, where the new resonances occur at a frequency shift of $\pm g^{\text{FP}}$ (as defined in Eq.~(\ref{eq:definition_g2})) with respect to the bare atomic resonance, the cascaded model predicts an additional shift, where both peaks of the vacuum-Rabi spectrum are shifted in the same direction, see Fig.~\ref{fig:beta2} in the main manuscript. The maximum shift $\pm\beta\gamma/2$ is reached for atoms close to one of the two cavity mirrors ($x_a=0,L/2$) and for $\alpha_0=\pi/2,3\pi/2,...$ (half-intensity points of the standing wave). Surprisingly, this shift is completely independent of the atom-resonator coupling strength or any cavity parameters and is just defined by $\beta$ and the longitudinal position $x_a$ of the atom in the cavity. 

The fact that one observes a total shift of the vacuum-Rabi splitting of $\beta\gamma\cdot \sin\alpha_0$ between the situations where an atom is placed close to mirror one ($x_a=0$) and two ($x_a=L/2$) can be understood in a semiclassical way, when considering the frequency-dependence of $\alpha$ in Eq.~(\ref{eq:alpha}). In order to measure the vacuum-Rabi splitting on has to perform a frequency scan of the coupled atom-resonator system. As a consequence, the standing wave pattern inside the cavity and thus the value of $\alpha$  will change with the frequency of the probe laser. When the atom is initially at a position of strong longitudinal intensity gradient ($\alpha_0=\pi/2,3\pi/2,...$) this leads to a change of the atom-resonator coupling strength and thus to a shift of the resonances of the coupled system by the same amplitude as given by Eq.~(\ref{eq:gshift}). 

\section{Beta factor in a Fabry-P\'erot resonator}\label{app:waist}
In order to relate the $\beta$ factor in our notation to the waist of the beam-mode in standard Fabry-Pérot resonators, one has to calculate the overlap of the Gaussian mode profile with the radiation pattern of the emitter which is given by the Hertzian dipole. Optimal coupling is achieved when the emitter is placed in the beam waist of the Gaussian mode. In this case, the mode of the Gaussian beam and the atom in the far field ($r\gg \lambda$) are given in spherical coordinates by
\begin{eqnarray}
\vec E_G&=&E_{0,G} \exp(ikr)\exp(-\frac{\theta^2}{\theta_0^2})\vec e_G(\theta,\phi) \\
\vec E_D&=&E_{0,D} \exp(ikr) (\vec e_r\times\vec e_D)\times \vec e_r,
\end{eqnarray}
where $\vec e_r$ is the unit vector in radial direction and $\vec e_G$ and $\vec e_D$ are the polarization unit vectors of the Gaussian mode and the dipole field, respectively. The factor
\begin{equation}
\theta_0=\frac{\lambda}{\pi w_0}
\end{equation}
is the divergence angle of the Gaussian mode with waist $w_0$. $E_{0,G}$ and $E_{0,D}$ are the amplitude of the two modes which we assume to be normalized such that
\begin{equation}
\int_{0}^{2\pi}\int_{0}^{\pi}d\phi \sin\theta d\theta |\vec E_G|^2=\int_{0}^{2\pi}\int_{0}^{\pi}d\phi \sin\theta d\theta |\vec E_D|^2=1.
\end{equation}
The probability $\beta$ that the emitter emits a photon into the Gaussian mode is then given by the overlap of the two field modes
\begin{equation}
\beta=\left|\int_{0}^{2\pi}\int_{0}^{\pi}d\phi \sin\theta d\theta \vec E_G\cdot\vec E_D^*\right|^2.
\end{equation} 

When the polarization of the Gaussian mode matches that of the dipole field $\vec e_G(0,\phi)=\vec e_D$, this yields, for not too small waists ($w_0\gtrapprox\lambda$), the analytic expression
\begin{equation}
\beta=\frac{3}{2\pi^2}\frac{\lambda^2}{w_0^2}.
\end{equation}

\end{document}